\documentclass{emulateapj}

\usepackage{graphicx}
\usepackage{graphics}

\def\gsim{\;\lower4pt\hbox{${\buildrel\displaystyle >\over\sim}$}\,}
\def\lsim{\;\lower4pt\hbox{${\buildrel\displaystyle <\over\sim}$}\,}

\slugcomment{}

\shorttitle{Modeling SNR Cassiopeia A from the SN to its Current
Age}
\shortauthors{S. Orlando et al.}

\begin{document}

\newcommand\rs[1]{_\mathrm{#1}}
\newcommand\op[1]{{\bf #1}}

\title{Modeling SNR Cassiopeia A from the Supernova Explosion to its
   Current Age:\\ The role of post-explosion anisotropies of ejecta}

\author{S. Orlando\altaffilmark{1}, M. Miceli\altaffilmark{2,1},
        M.L. Pumo\altaffilmark{1},
        F. Bocchino\altaffilmark{1}}
\email{orlando@astropa.inaf.it}

\altaffiltext{1}{INAF - Osservatorio Astronomico di Palermo ``G.S.
                 Vaiana'', Piazza del Parlamento 1, 90134 Palermo, Italy}
\altaffiltext{2}{Dip. di Fisica e Chimica, Univ. di Palermo, Piazza del
                 Parlamento 1, 90134 Palermo, Italy}

\begin{abstract}
The remnants of core-collapse supernovae (SNe) have complex
morphologies that may reflect asymmetries and structures developed
during the progenitor SN explosion. Here we investigate how the
morphology of the SNR Cassiopeia A (Cas\,A) reflects the characteristics
of the progenitor SN with the aim to derive the energies and masses
of the post-explosion anisotropies responsible for the observed
spatial distribution of Fe and Si/S.  We model the evolution of
Cas\,A from the immediate aftermath of the progenitor SN to the
three-dimensional interaction of the remnant with the surrounding
medium. The post-explosion structure of the ejecta is described by
small-scale clumping of material and larger-scale anisotropies. The
hydrodynamic multi-species simulations consider an appropriate
post-explosion isotopic composition of the ejecta. The observed
average expansion rate and shock velocities can be well reproduced
by models with ejecta mass $M_{\rm ej}\approx 4M_{\odot}$ and
explosion energy $E_{\rm SN}\approx 2.3\times 10^{51}$~erg. The
post-explosion anisotropies (pistons) reproduce the observed
distributions of Fe and Si/S if they had a total mass of $\approx
0.25\,M_{\odot}$ and a total kinetic energy of $\approx 1.5\times
10^{50}$~erg. The pistons produce a spatial inversion of ejecta
layers at the epoch of Cas\,A, leading to the Si/S-rich ejecta
physically interior to the Fe-rich ejecta. The pistons are also
responsible for the development of bright rings of Si/S-rich material
which form at the intersection between the reverse shock and the
material accumulated around the pistons during their propagation.
Our result supports the idea that the bulk of asymmetries observed
in Cas\,A are intrinsic to the explosion.
\end{abstract}


%

\section{Introduction}

It is generally accepted that the highly non-uniform distribution
of ejecta observed in core-collapse supernova remnants (SNRs) might
reflect pristine structures and features of the progenitor supernova
(SN) explosion (e.g. \citealt{2011ApJ...732..114L}). Thus the analysis
of inhomogeneities observed in the morphology of SNRs might help
to trace back the characteristics of the asymmetries that may have
occurred during the SN explosion, providing a physical insight into
the processes governing the SN engines. On the other hand, the
morphology of SNRs is expected to reflect also the interaction of the SN
blasts with the inhomogeneous ambient medium. Disentangling the
effect of this interaction from the effects of the SN explosion is
one of the major problems in linking the present day morphology of
SNRs to their SN progenitors.

The SNe-SNRs connection can be best studied in young SNRs, for which
the imprint of SN explosion on their morphology might be identified
more easily, before the remnants start to interact with the inhomogeneous
interstellar medium (ISM). Cassiopeia A (in the following Cas\,A)
is an attractive laboratory for studying the early evolutionary
phase of a SNR. In fact the observations suggest that its morphology and
expansion rate are consistent with the model of a remnant expanding
through the wind of the progenitor red supergiant (RSG) (e.g.
\citealt{2003ApJ...593L..23C, 2003ApJ...597..347L, 2009ApJ...703..883H,
2014ApJ...789....7L}).

Also Cas\,A is one of the best studied remnant and its three-dimensional
(3D) structure has been characterized in good details (e.g.
\citealt{2010ApJ...725.2038D, 2013ApJ...772..134M, 2015Sci...347..526M}).
One of the outstanding characteristics of its morphology is the
overall clumpiness, most likely due to pristine ejecta clumpiness
resulting from instabilities and mixing throughout the remnant
evolution. The masses of X-ray-emitting ejecta and the mass
distribution of various elements over the remnant have been derived
accurately from the observations (e.g. \citealt{2012ApJ...746..130H}).

From the 3D reconstruction of the spatial distribution of ejecta,
\cite{2010ApJ...725.2038D} suggested that the structure of Cas\,A
consists of a spherical component (roughly coincident with the
forward shock), a tilted thick disk, and several ejecta jets/protrusions,
the most prominent of which are the southeast (SE) and northwest
(NW) Fe-rich regions and the high velocity northeast (NE) and
southwest (SW) streams of Si-rich debris (often referred to as
``jets''). The jet/protrusion features have been interpreted as the
result of ``pistons'' of faster than average ejecta emerging from
the SN explosion (\citealt{2010ApJ...725.2038D}). According to this
scenario, the bright rings of ejecta clearly visible in Cas\,A and
circling the jets/pistons represent the intersection of these pistons
with the reverse shock.

An alternative explanation of the rings has been proposed by
\cite{2001ApJ...557..782B} who suggested that they represent
cross-sections of large cavities in the expanding ejecta created
by expanding plumes of radioactive $^{56}$Ni-rich ejecta (see also
\citealt{1993ApJ...419..824L}). This scenario is supported by the
analysis of near-infrared observations of Cas\,A that revealed a
bubble-like morphology of the remnant's interior that may originate
from the compression of surrounding non-radioactive material by the
expanding radioactive $^{56}$Ni-rich ejecta
(\citealt{2015Sci...347..526M}). Against this idea there is however
the advanced ionization age relative to other elements of X-ray
emitting shocked Fe (e.g. \citealt{2012ApJ...746..130H}). In fact
Fe-rich ejecta associated with the Ni bubble effect are expected
to have low ionization ages, at odds with observations.

A better understanding of the present day structure and chemical
stratification of ejecta in Cas\,A requires therefore to study the
evolution of chemically homogeneous ejecta layers since the SN
event, in order to map the layers at the explosion to the resulting
abundance pattern observed at the current age. Some effort in this
direction has been done for other SNRs mostly by using a one-dimensional
(1D) approach to overcome the difficulty of the very different time
and space scales of SNe and SNRs (e.g. \citealt{2008ApJ...680.1149B,
2014ApJ...785L..27Y, 2015ApJ...803..101P}). However these models
miss all the complex spatial structures (requiring 3D simulations)
observed in SNRs and so difficult to interpret. Recently a 3D model
describing the evolution of SN\,1987A since the SN event has been
used to identify the imprint of the SN on the remnant emission
(\citealt{2015ApJ...810..168O}).

In the attempt to link the ejecta structure of Cas\,A to the
properties of its progenitor SN, we developed a hydrodynamic model
describing the evolution of Cas\,A from the immediate aftermath of the
progenitor SN explosion, to the interaction of the remnant with the
RSG wind. The model considers complete and realistic conditions
of the early post-explosion ejecta structure, including the isotopic
composition of the ejecta appropriate for the expected progenitor
star.

In this paper we challenge the scenario of high velocity pistons
of ejecta emerging from the SN explosion proposed by
\cite{2010ApJ...725.2038D}. We describe the post-explosion
structure of the ejecta through small-scale clumping of material
and larger-scale anisotropies (possibly due to hydrodynamic
instabilities; e.g. \citealt{2006A&A...453..661K, 2008ARA&A..46..433W,
2010A&A...521A..38G}). We investigate the effects of the initial
ejecta structure on the final remnant morphology with the aim to
determine the energies and masses of the post-explosion anisotropies
responsible for the spatial distribution of Fe and Si/S observed
today in Cas\,A.

The paper is organized as follows. In Section~\ref{sec2} we describe
the hydrodynamic model and the numerical setup, in Section~\ref{sec3}
we discuss the results and, finally, we draw our conclusions in
Section~\ref{sec4}.

\section{Problem description and numerical setup}
\label{sec2}

Our simulations assume a pre-SN environment and initial conditions
for the SN explosion that are appropriate for a progenitor RSG (see
Sect.~\ref{prog_star}). Our approach follows that described by
\cite{2015ApJ...810..168O}: first, we simulate the post-explosion evolution
of the SN soon after the core-collapse (see Sect.~\ref{mod_sn});
then the output of these simulations are used to start 3D hydrodynamic
simulations describing the expansion of ejecta through the pre-SN
environment (see Sect.~\ref{mod_snr}).

\newpage
\subsection{The adopted progenitor star}
\label{prog_star}

The optical spectrum of the progenitor SN of Cas\,A (derived from
its scattered light echo; \citealt{2008Sci...320.1195K}) is remarkably
similar to that of the prototypical type IIb SN 1993J
(\citealt{1993Natur.364..507N}). Thus, as for SN 1993J, Cas\,A might
have originated from the collapse of a RSG with a 
main sequence (MS) mass of 13 to 20 $M_{\odot}$ that had lost
most of its hydrogen envelope before exploding
(\citealt{1993Natur.364..507N, 1994AJ....107..662A}).  This scenario
is also supported by several observational constraints that indicate
that the total ejecta mass of Cas\,A was only of 2 to 4 $M_{\odot}$
(\citealt{2006ApJ...640..891Y} and references therein) with about
$1-2\,M_{\odot}$ of oxygen (\citealt{2003A&A...398.1021W}). Considering
the presence of the neutron star, this ejecta mass is consistent
with a core mass at the end of the RSG phase of about 6
$M_{\odot}$ (as that inferred for SN 1993J; \citealt{1993Natur.364..507N});
the presence of a significant fraction of oxygen-rich ejecta suggests
a MS mass for the progenitor close to 20 $M_{\odot}$
(\citealt{1996ApJ...460..408T}).

The scenario of a progenitor with a MS mass of 20 $M_{\odot}$ poses
the problem if it evolved also through a Wolf-Rayet (WR)
phase\footnote{The lower limit of the mass of a MS star that can
evolve through a WR phase depends on the initial metallicity of the
star (e.g.  \citealt{2007ARA&A..45..177C}) and is considered to be
$> 20 M_{\odot}$ (\citealt{2005A&A...429..581M}).} or not. Several
studies suggest that the current morphology and expansion rate of
Cas\,A are consistent with a remnant interacting with the wind of
a RSG (e.g.  \citealt{2003ApJ...593L..23C, 2003ApJ...597..347L,
2009ApJ...703..883H, 2014ApJ...789....7L}). The presence of slow-moving
shocked circumstellar clumps in the remnant (the so-called
quasi-stationary flocculi) has been interpreted as signature of a
WR phase of the progenitor: the flocculi are fragments of the RSG
shell swept-up by a later WR wind (e.g. \citealt{1996A&A...316..133G}).
However, more recent studies have shown that these flocculi can be
explained as dense clumps in the RSG wind (\citealt{2003ApJ...593L..23C,
2009A&A...503..495V}) and that the morphology of Cas\,A is consistent
with an evolution of the progenitor without (or with a short - a
few thousand years) WR phase (\citealt{2008ApJ...686..399S,
2009A&A...503..495V}).

On the other hand, we note that the analysis of optical observations
suggests that stars with a MS mass above\footnote{However it should
be noted that this value is still questioned from both the theoretical
and the observational point of view (e.g. \citealt{2012ApJ...759...20K},
and references therein).} $\approx 17\,M_{\odot}$ should not appear
to explode as RSGs leading to standard type II Plateau SNe
(\citealt{2009ARAA..47...63S}). A similar result has been obtained
from the analysis of X-ray observations which put as upper limit
to the mass of a RSG star exploding as SN $M = 19\, M_{\odot}$
(\citealt{2014MNRAS.440.1917D}).  In the light of these considerations,
and following \cite{2009A&A...503..495V}, we assume for our simulations
that the progenitor of Cas\,A was a star with an initial mass between
15 and $20\,M_{\rm \odot}$ (according to the values suggested, e.g.
by \citealt{1994AJ....107..662A, 2014ApJ...789....7L}) that evolved
through a RSG phase and did not have a WR phase. Thus the pre-SN
environment immediately close to the SN was determined by the dense
slow wind from the RSG.

X-ray observations show that currently the remnant is still interacting
with this wind with a post-shock density ranging between 3 and 5
cm$^{-3}$ at the current outer radius of the remnant, $r_{\rm fs}
\approx 2.5$ pc (asuming a distance of $\approx 3.4$~kpc;
\citealt{2014ApJ...789....7L}). Neglecting the back-reaction of
accelerated cosmic rays (CRs) at the shock front, the upper limit
to the wind density at $r_{\rm fs} \approx 2.5$ pc is $n_{\rm w}
\approx 0.9\pm 0.3$~cm$^{-3}$. Assuming that the gas density in the
wind is proportional to $r^{-2}$ (where $r$ is the radial distance
from the progenitor), the amount of mass of the wind within the
radius of the forward shock $r_{\rm fs}$ is (\citealt{2014ApJ...789....7L})

\begin{equation}
\begin{array}{ll}
\displaystyle
M_{\rm w, sh} & = \int \rho(r)\, 4 \pi r^2 {\rm d}r = \int \rho_{\rm w} (r^2_{\rm
fs}/r^2)\, 4 \pi r^2 {\rm d}r \\ \\ \displaystyle
 & \approx 4 \pi \rho_{\rm w} r^3_{\rm fs} \approx 6 M_{\odot}
\end{array}
\label{wind_mass}
\end{equation}

\noindent 
where $\rho_{\rm w} = \mu m_{\rm H} n_{\rm w}$ is the mass density
of the wind at $r = r_{\rm fs}$, $\mu = 1.3$ is the mean atomic
mass (assuming cosmic abundances), and $m_H$ is the mass of the
hydrogen atom. Since the total mass lost during the RSG phase is
expected to be\footnote{This is the difference between the stellar
mass at the end of the MS phase, in the range between 15 and 20
$M_{\odot}$, and the core mass at the end of the RSG phase, 6
$M_{\odot}$.} between 9 and $14\,M_{\odot}$, we expect that the RSG
wind shell has a radius larger than $r_{\rm fs}$. Therefore, we can
safely assume that Cas\,A is still evolving through the RSG wind
with an initial $r^{-2}$ density profile.

\subsection{Modeling the post-explosion evolution of the supernova}
\label{mod_sn}

We modeled the post-explosion evolution of the SN by adopting a
1D Lagrangian code in spherical geometry. The code solves the equations
of relativistic radiation hydrodynamics, for a self-gravitating
matter fluid interacting with radiation, as described in detail in
\citet{2011ApJ...741...41P}. The code is fully general relativistic and
provides an accurate treatment of radiative transfer at all regimes, thus
allowing us to deal with optically thick and optically thin ejecta. The
code includes the coupling of the radiation moment equations with the
equations of relativistic hydrodynamics (during all the post-explosive
phases) and the heating effects associated with the decays of the
radioactive yields of the explosive nuclesynthesis process. Also, the
gravitational effects of the central compact object on the evolution of
the ejecta are taken into account.

We followed the evolution of the stellar ejecta from the shock
breakout at the stellar surface up to when the envelope has recombined
and the radioactive decays of the explosive nucleosynthesis products
dominates the energy budget (the so-called nebular stage). Also the
code computes the fallback of material on the central compact object
and consequently determines the amount of $^{56}$Ni in the ejected
envelope at late times. The code has been widely used to model the
observations of core-collapse SNe (e.g. \citealt{2014MNRAS.439.2873S,
2014MNRAS.438..368T,2014ApJ...787..139D,2012A&A...537A.141P}), having
also the capability of simulating the bolometric lightcurve and the time
evolution of the photospheric velocity and temperature.

We set the initial conditions of our simulations to mimic the
physical properties of the ejected material after shock passage following
the core-collapse (see \citealt{2011ApJ...741...41P} for details). The
parameters of our model setups are: the progenitor radius $R_0$, the total
ejecta energy $E_{\rm SN}$, the envelope mass at shock breakout $M_{\rm
env}$, and the total amount of $^{56}$Ni initially present in the ejected
envelope $M_{\rm Ni}$. Note that the value $M_{\rm env}$ indicates the
initial mass of the material surrounding the compact object (which has
a mass $M_{\rm cut}=1.6\,M_{\odot}$ at the onset of our simulations).
In all the simulations presented here, most of this mass is ejected in
the post-explosive phases, and only a minor part (of the order of a few
hundredths of a solar mass) falls back to the central object. We can then
conclude that the mass of the ejecta $M_{\rm ej}\approx M_{\rm env}$.

We explored different values of $R_0$, and $E_{\rm SN}$,
fixing the envelope mass $M_{\rm env}=4 M_{\odot}$ (according to
\citealt{2006ApJ...640..891Y} and \citealt{2009A&A...503..495V}), and
the initial amount of $^{56}$Ni $M_{\rm Ni} = 0.1 M_{\odot}$ (consistent
with the range of values derived by \citealt{2009ApJ...697...29E} for
Cas\,A\footnote{The adopted value is also consistent with the amount
of Ni synthesized during the explosion of a RSG star with a MS mass
between 15 and 20 $M_{\odot}$ (\citealt{1996ApJ...460..408T}).}).
In particular, we considered models with $E_{\rm SN}$ ranging
between 1 and $3\times 10^{51}$~erg, and $R_0$ ranging between 100
and 1000~$R_{\odot}$ (namely the range of values expected for a RSG;
e.g. \citealt{2005ApJ...628..973L}). In the set of models explored, the
maximum velocity of the ejecta immediately after the shock breakout is
in the range $8000-15000$~km~s$^{-1}$. For a given density of the RSG
wind, models with different values of $E_{\rm SN}$ and $R_0$ produce
remnants at $t = 340$~yr characterized by different radii and velocities
of the forward and reverse shocks. As explained in more details in
Sect.~\ref{mod_spher}, we searched for the values of $E_{\rm SN}$
and $R_0$ best reproducing altogether the density of the shocked RSG
wind inferred from observations (\citealt{2014ApJ...789....7L}) and the
radii and velocities of the forward and reverse shocks as observed at
the current time ($t \approx 340$~yr; see Table~\ref{tabobs}).

\begin{table}
\caption{Radii and Velocities of the Forward and Reverse Shocks
observed in Cas~A at the current age.}
\label{tabobs}
\begin{center}
\begin{tabular}{llll}
\hline
\hline
              &  Value                       & Reference \\
\hline
$r_{\rm FS}$  &  $\begin{array}{lll}2.'55 & \pm & 0.'2\\2.52 & \pm & 0.2 ~{\rm pc} \end{array}$    &  \cite{2001ApJ...552L..39G} \\
$r_{\rm RS}$  &  $\begin{array}{lll}1.'58 & \pm & 0.'16\\1.58 & \pm & 0.16 ~{\rm pc} \end{array}$  &  \cite{2001ApJ...552L..39G} \\
$v_{\rm FS}$  &  $4990 \pm 150$ km s$^{-1}$  & \cite{Vink1998} \\
              &                              & \cite{2004ApJ...613..343D} \\
$v_{\rm RS}^a$  &  $2000 \pm 400$ km s$^{-1}$  & \cite{2004ApJ...614..727M} \\
\hline
\end{tabular}
\end{center}
$^a$ Note that we consider the velocity of the reverse shock in the
observers frame (see \citealt{2009A&A...503..495V}).
\end{table}

\subsection{Modeling the evolution of the supernova remnant}
\label{mod_snr}

After we have simulated in 1D the post-explosion evolution of the SN
soon after the core-collapse (see Sect.~\ref{mod_sn}), we mapped
the output of these simulations in 3D and, then, started 3D hydrodynamic
simulations which describe the interaction of the remnant with the wind
of the progenitor RSG. We modeled the evolution of the blast wave by
numerically solving the time-dependent fluid equations of mass, momentum,
and energy conservation in a 3D Cartesian coordinate system $(x,y,z)$;
the hydrodynamic equations were extended to include the effects of the
radiative losses from an optically thin plasma:

\begin{equation}
\begin{array}{l}\displaystyle
\frac{\partial \rho}{\partial t} + \nabla \cdot \rho \mbox{\bf u} = 0~,
\\ \\ \displaystyle
\frac{\partial \rho \mbox{\bf u}}{\partial t} + \nabla \cdot \rho
\mbox{\bf uu} + \nabla P = 0~,
\\ \\ \displaystyle
\frac{\partial \rho E}{\partial t} +\nabla\cdot (\rho E+P)\mbox{\bf u}
= -n_{\rm e} n_{\rm H} \Lambda(T)~,
\end{array}
\label{mod_eq}
\end{equation}

\begin{figure}[!t]
  \centering \includegraphics[width=8.5cm]{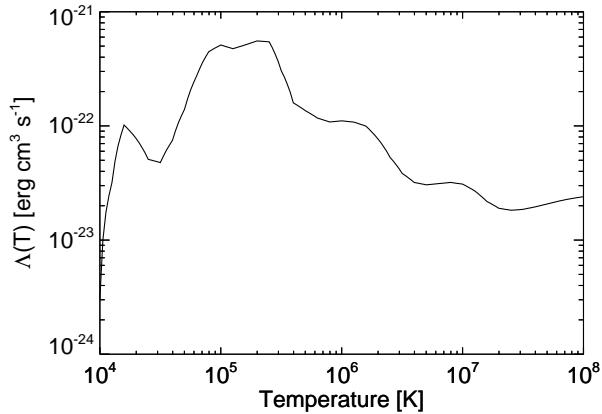}
  \caption{Radiative losses for an optically thin plasma assuming the
           cosmic metal abundances (e.g. \citealt{mgv85,
           2000adnx.conf..161K}).}
  \label{rad_loss}
\end{figure}

\bigskip
\noindent
where $E = \epsilon + |\mbox{\bf u}|^2/2$ is the total gas energy
(internal energy, $\epsilon$, and kinetic energy), $t$ is the time,
$\rho = \mu m_H n_{\rm H}$ is the mass density, $\mu$ is the mean
atomic mass which depends on the radial distance (as explained below),
$n_{\rm H}$ is the hydrogen number density, $n_{\rm e}$ is the
electron number density, {\bf u} is the gas velocity, $T$ is the
temperature, and $\Lambda(T)$ represents the radiative losses per
unit emission measure (e.g. \citealt{mgv85, 2000adnx.conf..161K};
see Fig~\ref{rad_loss}). We used the ideal gas law,
$P=(\gamma-1) \rho \epsilon$, where $\gamma$ is the adiabatic index.

We used the FLASH code (\citealt{for00}) to perform the
calculations. In particular we solved the equations for compressible gas
dynamics with the FLASH implementation of the piecewise-parabolic method
(\citealt{cole84}). The radiative losses $\Lambda$ in Eq.~\ref{mod_eq}
are calculated through a table lookup/interpolation method. Also we
extended the code by additional computational modules to calculate
the deviations from electron-proton temperature-equilibration and
the deviations from equilibrium of ionization of the most abundant
ions. For the former, we calculated the ion and electron temperatures in
each cell of the post-shock medium, taking into account the effects of
Coulomb collisions (see \citealt{2015ApJ...810..168O} for the details
of the implementation). According to \cite{2007ApJ...654L..69G},
first the electrons are assumed to be heated at the shock front almost
istantaneously up to $kT \sim 0.3$~keV by lower hybrid waves. This
istantaneous heating does not depend on the shock Mach number and is
expected for fast shocks (i.e. $> 10^3$ km s$^{-1}$) as those simulated
here. Then we considered the effects of the Coulomb collisions to
calculate the evolution of ion and electron temperatures in each cell of
the post-shock medium in the time $\Delta t_{\rm j} = t-t_{\rm shj}$,
where $t_{\rm shj}$ is the time when the plasma in the $j$-th domain cell
was shocked and $t$ is the current time. The time $t_{\rm shj}$ is stored
in an additional passive tracer added to the model equations. To estimate
the deviations from equilibrium of ionization of the most abundant ions,
we adopted the approach suggested by \cite{2010MNRAS.407..812D}. In fact,
this approach ensures high efficiency in the calculation (expecially
in the case of 3D simulations as in our case) together with a reasonable
accuracy in the evaluation of non-equilibrium of ionization effects.
The approach consists in the computation of the maximum ionization
age in each cell of the spatial domain $\tau_{\rm j} = n_{\rm ej} \Delta
t_{\rm j}$, where $n_{\rm ej}$ is the electron density in the
j-th cell and $\Delta t_{\rm j}$ is the time since when the plasma in
the cell was shocked (see above).

\begin{figure}[!t]
  \centering \includegraphics[width=8.5cm]{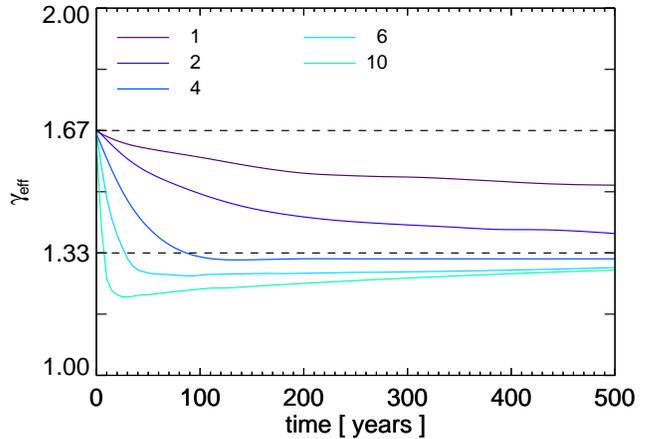}
  \caption{Evolution of the effective adiabatic index $\gamma\rs{eff}$
  at the shock front for different values of the injection rate
  $\eta$ (see legend in units of $10^{-4}$) (derived from
  \citealt{2010A&A...509L..10F}).}
  \label{eff_gamma}
\end{figure}

The non-thermal emission detected in Cas\,A indicates that effective
acceleration of CRs to energies exceeding 100 TeV occurs at the
shock fronts (e.g. \citealt{2010ApJ...710L..92A, 2013ApJ...779..117Y}).
Thus we included in the model also the modifications of the shock
dynamics due to the back-reaction of accelerated CRs by
following the approach of \citet{2010A&A...509L..10F} (see also
\citealt{2012ApJ...749..156O}). More in details, we included an effective
adiabatic index $\gamma\rs{eff}$ which depends on the injection rate of
particles $\eta$ (i.e. the fraction of ISM particles entering the shock
front) and on the time. At each time-step of integration,
$\gamma\rs{eff}$ is calculated at the shock front through linear
interpolation from a lookup table derived by \citet{2010A&A...509L..10F}
on the basis of the semi-analytical model of \cite{2002APh....16..429B,
2004APh....21...45B}. Then $\gamma\rs{eff}$ is advected within the
remnant, remaining constant in each fluid element. Fig.~\ref{eff_gamma}
shows the effective adiabatic index $\gamma\rs{eff}$ at the shock front
versus time for constant values of the injection rate $\eta$ (derived
from \citealt{2010A&A...509L..10F}).

We started the 3D simulations once almost all ($> 98$\%) the
explosion energy is kinetic (in most of our simulations this happens
few days after the SN explosion). Then we followed the expansion
of the remnant through the RSG wind during the first 340 years of
evolution. The initial remnant radius varies between 20 and 200~AU
(namely between $10^{-4}$ and $10^{-3}$~pc), depending on the initial
radius of the progenitor star. The output of the SN simulations
provides the initial radial structure of the ejecta for the SNR
simulations.

Several theoretical studies predict that the remnants of core-collapse
SNe are characterized by small-scale clumping of material and
larger-scale anisotropies (e.g. \citealt{1993ApJ...419..824L,
2002ApJ...579..671W, 2006A&A...453..661K, 2008ARA&A..46..433W,
2010A&A...521A..38G}, and references therein). Since these structures
cannot be described by our 1D SN simulations, we account for them
by prescribing an initial clumpy structure of the ejecta and
large-scale anisotropies (as suggested by \citealt{2006A&A...453..661K}),
after the 1D radial density distribution of ejecta (calculated with
the SN simulations) is mapped into 3D.

The small-scale ejecta clumps are modeled as per-cell random
density perturbations\footnote{We define density perturbation the
density contrast of the clump with respect to the density in the region
occupied by the clump if the perturbation was not present.}. Following
\cite{2012ApJ...749..156O}, we derived these perturbations by adopting a
power-law probability distribution with index $\xi = -1$. The parameter
characterizing the distribution is the maximum density perturbation
that is possible to reach in the simulation. For the purposes of this
paper, we assumed that the small-scale clumps have initial size about
2\% of the initial remnant radius and a maximum density contrast 
(namely a density perturbation) $\nu_{\rm max} = 5$. These values are in
agreement with those suggested by spectropolarimetric studies of SNe (e.g.
\citealt{2003ApJ...591.1110W, 2004ApJ...604L..53W, 2008ARA&A..46..433W,
2010ApJ...720.1500H}).

The post-explosion large-scale anisotropies in the ejecta distribution
are modeled as overdense spherical knots (hereafter called ``shrapnels'')
in pressure equilibrium with the surrounding ejecta\footnote{Through
additional simulations, we checked that the results do not change
significantly if the initial temperature of the shrapnel was the same
as that of the surrounding ejecta.}. Our primary goal is to derive the
mass and energy of the anisotropies responsible for the inhomogeneous
distribution of Fe and Si/S observed today in Cas\,A. To this end, we
explored the space of parameters characterizing the initial shrapnels
to find those best reproducing the observations.  In particular,
we considered shrapnels initially located either within or outside
the iron core, at distance $D_{\rm knot}$ from the center, with
radius $r_{\rm knot}$ ranging between 3\% and 10\% of the initial
remnant radius, with density between 10 and 100 times larger than
those of the surrounding ejecta at distance $D_{\rm knot}$ (density
contrast $\chi_{\rm n}$), and with radial velocity between 1 and 3.5
times larger than that of the surrounding ejecta (velocity contrast
$\chi_{\rm v}$). These ranges of values are consistent with those
derived from multi-dimensional simulations of core-collapse SNe which
show that Rayleigh-Taylor instabilities are seeded by the flow-structures
resulting from neutrino-driven convection and are effective at creating
metal-rich knots at the terminal ends of Rayleigh-Taylor fingers
(e.g. \citealt{2003A&A...408..621K, 2012ApJ...755..160E}). These knots
present dimensions ranging between 2\% and 16\% of the remnant radius at
the time of knot formation and are more than one order of magnitude denser
than the surrounding ejecta (e.g.  \citealt{2012ApJ...755..160E}). Also
the simulations show that the metal fingers and clumps are much faster
than the surrounding medium (with velocities up to 5000~km s$^{-1}$)
and are correlated with the biggest and fastest-rising plumes of
neutrino-heated matter (e.g. \citealt{2015A&A...577A..48W}).

\begin{figure}[!t]
  \centering \includegraphics[width=8.5cm]{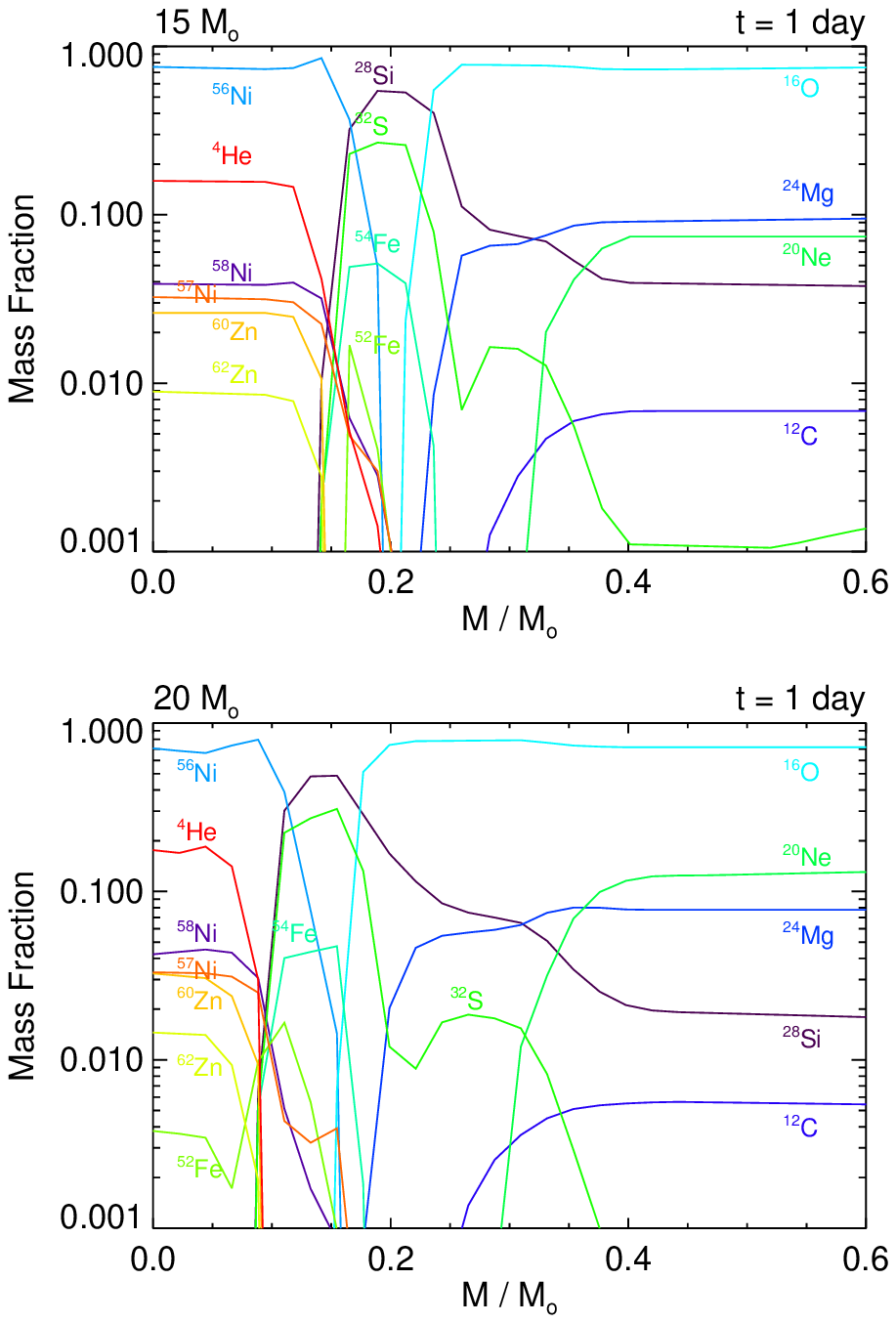}
  \caption{Isotopic composition of the ejecta as it results from
  post-shock SN processing of a core-collapse SN originating from
  a $15\,M_{\rm \odot}$ (upper panel) and $20\,M_{\rm \odot}$ (lower
  panel) MS star (derived from \citealt{1996ApJ...460..408T}). Only
  the dominant abundances of intermediate mass nuclei are considered.}
  \label{mass_fraction}
\end{figure}

As initial isotopic composition of the ejecta, we adopted that
derived by \cite{1996ApJ...460..408T} for a core-collapse SN either
from a $15\,M_{\rm \odot}$ or from a $20\,M_{\rm \odot}$ MS star,
namely the extremes of the range of masses suggested in the
literature (e.g. \citealt{1994AJ....107..662A, 2014ApJ...789....7L}).
Fig.~\ref{mass_fraction} shows the dominant abundances of elements 
that we follow in details during the evolution. These elements are
those that can allow us to compare the distribution of chemical
homogeneous regions of ejecta derived from the simulations with
those derived from the analysis of observations (e.g.
\citealt{2010ApJ...725.2038D, 2012ApJ...746..130H, 2013ApJ...772..134M,
2015Sci...347..526M}). The mean atomic mass $\mu$ for the ejecta
considers their isotopic composition (\citealt{1996ApJ...460..408T}),
whereas $\mu = 1.3$ for the RSG wind, assuming cosmic abundances.

As discussed in Sect.~\ref{prog_star}, we assume that the blast wave from
the SN explosion propagates through the wind of the progenitor RSG during
the whole evolution. The wind is assumed to be spherically symmetric
with gas density proportional to $r^{-2}$ (where $r$ is the radial
distance from the progenitor). The wind density at $r = 2.5$~pc is
constrained by X-ray observations of the shocked wind
(\citealt{2014ApJ...789....7L}). Since the shock compression
ratio varies with the injection rate, the wind density is different
for models with different $\eta$ (see Table~\ref{tabmod}) in order
to obtain the same post-shock density as inferred from the
observations.

\begin{table}
\caption{Parameters for the Models of Cas\,A best reproducing the
observations}
\label{tabmod}
\begin{center}
\begin{tabular}{lccc}
\hline
\hline
Model$^a$ &  $\eta$  &  $n_{\rm w}$  & $E_{\rm cr}$ \\
          &          &  [cm$^{-3}$]  & [$10^{50}$~erg] \\
\hline
SN-4M-2.3E-0ETA  &    0                  &  0.9   &  0     \\
SN-4M-2.3E-1ETA  &    $1\times 10^{-4}$  &  0.81  &  0.94  \\
SN-4M-2.3E-2ETA  &    $2\times 10^{-4}$  &  0.62  &  1.80  \\
SN-4M-2.3E-4ETA  &    $4\times 10^{-4}$  &  0.50  &  2.20  \\
SN-4M-2.3E-6ETA  &    $6\times 10^{-4}$  &  0.44  &  2.28  \\
SN-4M-2.3E-10ETA &    $1\times 10^{-3}$  &  0.42  &  2.34  \\
\hline
\end{tabular}
\end{center}
$^a$ In all these models, the explosion energy is $E_{\rm SN} =
2.3\times 10^{51}$~erg, the ejecta mass is $M_{\rm env} = 4
M_{\rm \odot}$, and the progenitor pre-SN radius is $R_0 =
350\,R_{\odot}$.
\end{table}

We studied the chemical evolution of the ejecta by adopting the
multiple fluids approach present in FLASH (\citealt{for00}). Each
fluid is associated to one of the heavy elements shown in
Fig.~\ref{mass_fraction} and initialized with the corresponding
abundances of elements reported in the figure. In such a way, we
followed the evolution of the isotopic composition of the ejecta
and mapped the spatial distribution of heavy elements at the present
epoch. During the remnant evolution, the different fluids mix
together. At any time $t$ the density of a specific element in a
fluid cell is given by $\rho_{\rm el} = \rho C_{\rm el}$, where
$C_{\rm el}$ is the mass fractions of each element and the
index ``el'' refers to a different element.

The SN explosion is assumed to sit at the origin of the
3D Cartesian coordinate system $(x_0, y_0, z_0) = (0, 0, 0)$. The
computational domain extends 6~pc in the $x$, $y$, and $z$ directions
(the current outer radius of the remnant is $\approx 2.5$~pc, assuming
a distance of $\approx 3.4$~kpc; \citealt{1995ApJ...440..706R}). We assume
zero-gradient (outflow) conditions at all boundaries.

The SN explosion and subsequent evolution of Cas\,A involve
very different length and time scales, from the relatively small size
of the very fast evolving system in the immediate aftermath of the SN
explosion (the remnant radius at the beginning of the 3D hydrodynamic
simulations is between $10^{-4}$ and $10^{-3}$~pc) to the larger extension
of the slowly expanding remnant (the final remnant radius is $\approx
2.5$~pc). This makes rather challenging the 3D modeling of Cas\,A and we
were able to capture the very different scales involved, by exploiting
the adaptive mesh refinement capabilities of FLASH. More specifically,
we employed 20 nested levels of refinement, with resolution increasing
twice at each refinement level. The refinement/derefinement of the mesh
is guided by the changes in mass density and temperature and follows
the criterion of \cite{loehner}. In addition we kept the
computational cost approximately constant during the evolution, by
adopting an automatic mesh derefinement scheme in the whole spatial
domain (\citealt{2012MNRAS.419.2329O}): we gradually decreased the
maximum number of refinement levels from 20 (at the beginning of the
simulations) to 6 (at the end) following the expansion of the remnant. The
effective spatial resolution reached at the finest level was $\approx
10^{-6}$~pc ($\approx 10^{-2}$~pc) at the beginning (at the end) of the
simulations, corresponding to an effective mesh size of $[5\times 10^6]^3$
($[512]^3$). In such a way, the number of grid zones per radius of
the remnant $N_{\rm pt}$ was $> 100$ during the whole evolution, with
$N_{\rm pt} \approx 100$ at $t=0$ and $N_{\rm pt} > 250$ at $t= 340$~yr.

\section{Results}
\label{sec3}

\subsection{The case of a spherically symmetric explosion} 
\label{mod_spher}

As a first step, we explored the parameter space of the SN-SNR
model, assuming a 3D spherically symmetric SN explosion. The
simulations considered, therefore, include the small-scale clumpy
structure of the ejecta but do not consider any large-scale anisotropy
(i.e. the shrapnels; see Sect.~\ref{mod_snr}). The aim was to
derive the best-fit basic parameters of the model (ejecta mass,
explosion energy, RSG wind density, and efficiency of CR acceleration)
by comparing model results with observations, in view of the study
concerning the effects of large-scale anisotropies on the remnant
morphology.

Our observing constraints are the radii and velocities of the forward
and reverse shocks as observed at current time (reported in
Table~\ref{tabobs}; see also \citealt{2009A&A...503..495V}). Another
constraint is the density of the shocked RSG wind that is inferred
to range between 3 and 5 cm$^{-3}$ from the analysis of {\it Chandra}
observations (\citealt{2014ApJ...789....7L}). Thus we searched for
the parameters ($R_0$, $E_{\rm SN}$, and $M_{\rm env}$) of the SN
model which reproduce altogether the observed density of the shocked
wind and the observed radii and velocities of the forward and reverse
shocks. Since the non-thermal emission detected in Cas\,A indicates
that effective acceleration of CRs occurs at the shock fronts (e.g.
\citealt{2010ApJ...710L..92A}), we explored also different values of
the injection rate $\eta$ (see Sect.~\ref{mod_snr}) to account the
feedback of CR acceleration on the remnant expansion. Considering
that the shock compression ratio increases with $\eta$ (due to a
faster decrease of the effective adiabatic index; see
Fig.~\ref{eff_gamma}), we varied accordingly the pre-shock wind
density $n_{\rm w}$ at $r=2.5$~pc (namely the current outer radius
of Cas A). 

In all the cases explored, the SN-SNR simulations consist of two
main phases: the post-explosion evolution of the supernova (lasting
few hundreds days since the outburst) and the transition from SN
to SNR. The first phase starts when the shock wave following the
core-collapse reaches the stellar surface. Then the evolution follows
the general trend described by \cite{2011ApJ...741...41P} and it
is briefly summarized below. Initially the envelope is completely
ionized and optically thick. Most of the internal energy is gradually
released, contributing to the gross emission of the SN. Few days after,
the ejecta start to recombine and the shock front recedes through
the envelope. In this phase, the resulting sudden release of energy
dominates the SN emission. After the envelope is fully recombined and
optically thin to optical photons, the SN emission originates from the
thermalization of the energy deposited by $\gamma$-ray photons.

The second phase starts few hundreds days after the SN event. We
followed the transition from SN to SNR for 340 years. During this
time a forward and a reverse shocks are formed, the former
propagating into the RSG wind and the latter driven back into the
ejecta. The ejecta clumps interact with each other and enhance the
development of hydrodynamic instabilities that enhance the mixing
of layers with different isotopic composition.

\begin{figure}[!t]
  \centering \includegraphics[width=8.5cm]{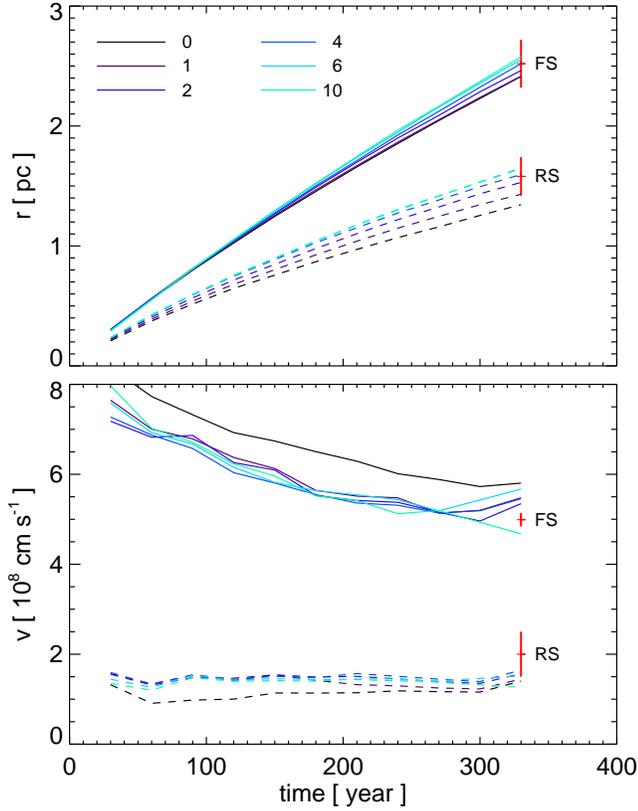}
  \caption{Angle-averaged radii (upper panel) and velocities (lower
  panel) of the forward (solid lines) and reverse (dashed lines)
  shocks vs. time for models assuming a different value of injection
  rate $\eta$ (see legend in units of $10^{-4}$; see also
  Fig.~\ref{eff_gamma}).  The red crosses show the corresponding
  observational values at the current age of Cas A (see Table
  \ref{tabobs}); the vertical lines of the crosses show the
  observational uncertainty.} \label{pos_shock}
\end{figure}

At $t = 340$~yr we compared the angle-averaged radii and velocities
of the forward and reverse shocks resulting from our models with
those observed. Fig.~\ref{pos_shock} shows these quantities for the
models best reproducing the observations together with the observed
values at the current epoch. The models in the figure differ from
each other for the injection rate $\eta$ and, consequently, for the
wind density $n_{\rm w}$ at $r=2.5$~pc (see Table~\ref{tabmod}).
We found that the observations are best reproduced by models
characterized by a total ejecta energy $E_{\rm SN} = 2.3\times
10^{51}$~erg, an envelope mass $M_{\rm env} = 4\,M_{\odot}$ (which
has been fixed in our simulations), and a progenitor radius
$R_0 = 350\,R_{\odot}$. Our best-fit explosion energy is in good
agreement with the value inferred from the observations, $E_{\rm
SN} \approx 2\times 10^{51}$~erg (e.g. \citealt{2003ApJ...597..347L,
2003ApJ...597..362H}); the ejecta mass is within the range of
values discussed in the literature, $M_{\rm ej} = 2-4M_{\odot}$
(e.g. \citealt{2003ApJ...597..347L, 2003ApJ...597..362H,
2006ApJ...640..891Y}). We note that our model predicts a total
energy which is smaller than that found by \cite{2003ApJ...593L..23C},
$E_{\rm SN} \approx 4\times 10^{51}$~erg, on the basis of a 1D
hydrodynamic model. Apart from the effects of ejecta clumping that are
not included in their model, these authors consider a mass of ejecta,
$M_{\rm ej} = 3.2\,M_{\odot}$, which is significantly smaller than
that adopted here. As a consequence, in their case, a larger total
explosion energy is required to fit the observed radius of Cas\,A.
Table~\ref{tabmod} summarizes the basic parameters characterizing the
SN-SNR models which best-fit our observing constraints: injection rate
$\eta$, wind density $n_{\rm w}$ at $r = 2.5$~pc, and the fraction of
explosion energy converted to CRs $E_{\rm cr}$ at $t = 340$~yr; all
these models have the same ejecta mass $M_{\rm env}$, explosion energy
$E_{\rm SN}$, and progenitor pre-SN radius $R_0$.

\begin{figure}[!t]
  \centering \includegraphics[width=9.0cm]{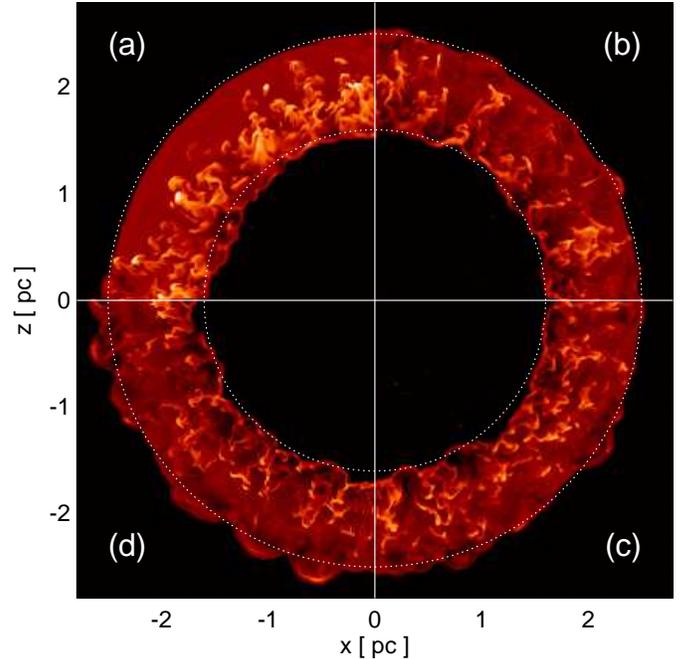}
  \caption{2D sections in the $(x,z)$ plane of the spatial density
  distribution of shocked plasma at $t=340$~yr for runs SN-4M-2.3E-1ETA
  (a), SN-4M-2.3E-4ETA (b), SN-4M-2.3E-6ETA (c), and SN-4M-2.3E-10ETA
  (d). The dotted circles mark the observed average position of the
  forward and reverse shocks in the $(x,z)$ plane.} \label{fig_injeff}
\end{figure}

As expected, the back-reaction of accelerated CRs mainly
affects the density structure of the region between the forward and
reverse shocks. There the plasma is characterized by an effective
adiabatic index $\gamma\rs{eff}$ which depends on $\eta$ and varies
between $\approx 4/3$ and $5/3$ (see Fig.~\ref{eff_gamma}). As a
result, the density jump at the shock $\sigma =
(\gamma\rs{eff}+1)/(\gamma\rs{eff}-1)$ varies between 7 and 4. The
ejecta clumping enhances the growth of Rayleigh-Taylor instabilities
at the contact discontinuity (\citealt{2012ApJ...749..156O}). The
CR acceleration enhances even further these instabilities. This is
shown in Fig.~\ref{fig_injeff} that presents 2D sections in the
$(x,z)$ plane of the spatial distribution of plasma density at
$t=340$~yr for models with different $\eta$ (runs SN-4M-2.3E-1ETA,
SN-4M-2.3E-4ETA, SN-4M-2.3E-6ETA, and SN-4M-2.3E-10ETA). As a
consequence of the enhanced intershock Rayleigh-Taylor mixing, the
shell of shocked wind is thinner at the forward shock for higher
values of $\eta$ and, consequently, the separation between the
forward shock and the contact discontinuity is smaller. Panel (c)
and, especially, panel (d) of Fig.~\ref{fig_injeff} also show that
the enhanced mixing can easily spread the ejecta material close to,
or even beyond, the average radius of the forward shock, depending
on the size and density contrast of the initial clumps (see also
\citealt{2012ApJ...749..156O, 2013MNRAS.430.2864M} for more details).

Fig.~\ref{pos_shock} shows that models differing for the injection
rate are all able to fit quite well the radius of the forward shock
within the observational uncertainty so that, from the comparison
of model results with observations, it is not possible to constrain
the value of $\eta$.  Nevertheless we note that models with low
values of $\eta$ tend to underestimate the radius of the reverse
shock and overestimate the velocity of the forward shock. In general
our models predict a velocity of the reverse shock which is slightly
lower than observed.

For each of the models fitting our observing constraints, we derived
the fraction of explosion energy converted to CRs, $E_{\rm cr}$, and
compared the simulated values (see Table~\ref{tabmod}) with those
inferred from observations. From the analysis of {\it Fermi} data,
\cite{2010ApJ...710L..92A} estimated the total content of accelerated CRs
as $\approx (1-4) \times 10^{49}$ erg. A similar result has been found
by \cite{2013ApJ...779..117Y} who suggest that the total energy lost
amounts to $\approx 4 \times 10^{49}$ erg\footnote{It is worth noting
that \cite{2014ApJ...785..130Z} suggest an energy lost closer to $\approx
3 \times 10^{50}$ erg in Cas\,A, on the base of the results of a model
describing the diffusive shock acceleration of particles in the non-linear
regime.}. In our model, $E_{\rm cr}$ increases for higher values of
$\eta$ and ranges between $9\times 10^{49}$~erg ($\eta = 10^{-4}$) and
$2.3\times 10^{50}$~erg ($\eta = 10^{-3}$), higher than those inferred
from the observations. We conclude therefore that the injection rate
in Cas\,A should be slightly lower than $\eta = 10^{-4}$.

\begin{figure}[!t]
  \centering \includegraphics[width=8.5cm]{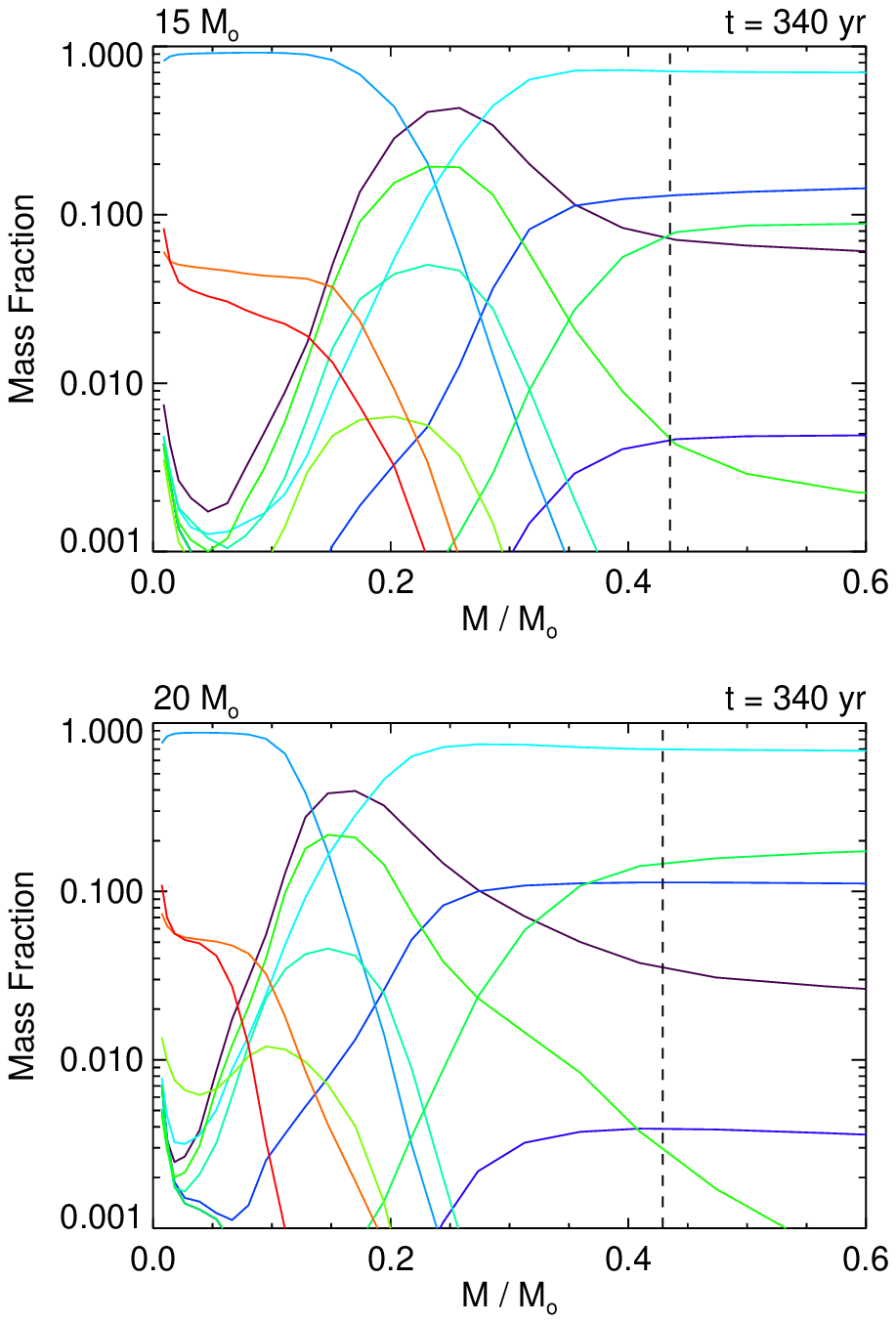}
  \caption{As in Fig.~\ref{mass_fraction} at $t=340$~yr for run
  SN-4M-2.3E-1ETA. The dashed lines show the position of the reverse
  shock.}
  \label{evolved_chemical}
\end{figure}

During the remnant expansion, we followed the evolution of the
isotopic composition of ejecta, focusing on the fluids tracing the
isotopes of Fe, Si, and S (see Fig.~\ref{evolved_chemical}), namely
those characterizing most of the anisotropies (e.g. jets, pistons)
identified in the morphology of Cas\,A (e.g.
\citealt{2010ApJ...725.2038D}). We investigated their spatial
distribution at $t=340$~yr in the case of a progenitor MS star of
either $15 M_{\odot}$ or $20 M_{\odot}$ and estimated the fraction
of their mass which is expected to be shocked. We found that, in
average, the stratification of chemical layers at the present age
reflects the radial distribution of ejecta in the immediate aftermath
of the progenitor SN. It is interesting to note that, from observations,
there is evidence that the onion-skin nucleosynthetic layering of
the SN has been preserved in some regions of Cas\,A
(\citealt{2006ApJ...645..283F, 2010ApJ...725.2038D}).

On the other hand, in all models of spherically symmetric explosion,
the masses of shocked Si and S are significantly lower (by more
than 40\%) than inferred from observations. More important, we found
that no significant amount ($< 10^{-3} M_{\odot}$) of shocked Fe
is predicted at odds with the results of observations
(\citealt{2012ApJ...746..130H}). At first sight, this result may
appear to be not surprising, given that the Fe lies deeply in the
remnant interior in all our models and the bulk of it has not been
reached by the reverse shock at the age of Cas\,A. However, the
ejecta are known to be characterized by a clumpy structure. This
may lead to the mixing of initially chemically homogeneous layers
and to some overturning of the ejecta due to hydrodynamic instabilities
developing during clumps interaction. Our 3D hydrodynamic model
shows that the small-scale clumping of ejecta cannot account for
the observed redistribution of elements in chemically distinct
layers of ejecta in Cas\,A.

\subsection{Effect of a post-explosion anisotropy of ejecta}
\label{sing_anis}

The results of Sect.~\ref{mod_spher} strongly support the idea that
violent dynamical processes have characterized the SN explosion and
led to a redistribution of heavy elements in the outer chemical
layers. Possible examples of these processes are uneven neutrino
heating, axisymmetric magnetorotational effects, hydrodynamic
instabilities (e.g. \citealt{2003A&A...408..621K, 2002ApJ...568..807W}).

Here we investigate how the parameters (size, density, and velocity)
of a post-explosion anisotropy formed few hours after the SN event
may determine the redistribution of Fe, Si, and S in the outer
chemical layers of the remnant. As discussed in Sect.~\ref{mod_snr},
the anisotropy is included in the simulations by modeling a large-scale
spherical knot (a shrapnel) with given size, density, and velocity.
To reduce the computational cost, we performed 3D simulations by
considering only one octant of the whole spatial domain (namely the
domain extends between 0 and 3 pc in the $x$, $y$, and $z$ directions;
see Sect~\ref{mod_snr}) and limiting the study to the case of a
progenitor MS star of $20\,M_{\odot}$. Reflecting boundary conditions
were used at $x=0$, $y=0$, and $z=0$, consistent with the adopted
symmetry. Note that analogous simulations have been performed by
\cite{2013MNRAS.430.2864M} and \cite{2015MNRAS.453..166T} but in
two-dimensions (2D). A major difference with those simulations is
that our 3D model describes the post-explosion evolution of the SN
and traces the evolution of the isotopic composition of ejecta since
few hours after the SN event.

For the basic parameters of the SN-SNR model (mass of ejecta, energy
of explosion, injection rate, density of the RSG wind, etc.), we
considered those of run SN-4M-2.3E-1ETA (see Table~\ref{tabmod})
which is one of the models reproducing the main observing
constraints\footnote{Namely the observed density of the shocked
wind and the observed average radii and velocities of the forward
and reverse shocks.} of Cas\,A. We preferred this model to the
others because it predicts a fraction of explosion energy converted
to CRs that is closer to the values inferred from observations (a
factor of $\approx 2$ larger than that suggested by
\citealt{2010ApJ...710L..92A} and \citealt{2013ApJ...779..117Y})

We explored the case of a shrapnel initially located either within
the iron core at the distance $D_{\rm knot} = 0.15\, R_{\rm SNR}$
from the center (where $R_{\rm SNR}\approx 10$~AU is the initial
radius of the remnant 1 day after the SN event) or just outside the
iron core at $D_{\rm knot} = 0.35\, R_{\rm SNR}$. In all the cases,
the shrapnel is placed at an angle of $45^o$ with respect to the
$x$, $y$, and $z$-axis. The initial radius of the shrapnel is
$r_{\rm knot}$, its density and velocity are $\chi_{\rm n}$ and
$\chi_{\rm v}$ times larger, respectively, than those of the
surrounding ejecta at distance $D_{\rm knot}$ from the center. A
summary of the cases explored is given in Table~\ref{tab_knot}.

\begin{table}
\caption{Summary of the models describing the evolution of a
post-explosion anisotropy of ejecta}
\label{tab_knot}
\begin{center}
\begin{tabular}{lcccc}
\hline
\hline
Model  &  $D_{\rm knot}$   &  $r_{\rm knot}$   &  $\chi_{\rm n}$  &  $\chi_{\rm v}$ \\
       &  [$R_{\rm SNR}$]  &  [$R_{\rm SNR}$]  &                  &                 \\
\hline
Fe-R6-D10-V3.5  &  0.15  &  0.06  &  10   &  3.5  \\
Fe-R6-D50-V3.5  &  0.15  &  0.06  &  50   &  3.5  \\
Fe-R6-D100-V3.5 &  0.15  &  0.06  &  100  &  3.5  \\
Fe-R4-D100-V1.5 &  0.15  &  0.04  &  100  &  1.5  \\
Fe-R4-D100-V2.5 &  0.15  &  0.04  &  100  &  2.5  \\
Fe-R4-D100-V3.5 &  0.15  &  0.04  &  100  &  3.5  \\
Fe-R3-D100-V3.5 &  0.15  &  0.03  &  100  &  3.5  \\
Si-R10-D10-V1 &  0.35  &  0.1  &  10  &  1  \\
Si-R10-D5-V3  &  0.35  &  0.1  &  5   &  3  \\
Si-R10-D3-V5  &  0.35  &  0.1  &  3   &  5  \\
Si-R10-D1-V10 &  0.35  &  0.1  &  1   &  10  \\
\hline
\end{tabular}
\end{center}
\end{table}

The evolution of the shrapnel is analogous to that described by
\cite{2013MNRAS.430.2864M}, except for the interaction with the
surrounding small-scale clumps of ejecta which are present in our
simulations.  Initially the shrapnel pushes out through less dense
and chemically distinct layers above, favoring the development of
hydrodynamic instabilities at its boundary which contribute to its
fragmentation. Then the shrapnel enters in the intershock region
where it is compressed, heated, and ionized. Also it interacts with
Rayleigh-Taylor and Richtmyer-Meshkov instabilities developed at
the contact discontinuity. As a result, the shrapnel is partially
eroded by the instabilities and evolves towards a core-plume
structure. At this stage, its core can be significantly denser than
the surrounding shocked ejecta, depending on the initial size and
density contrast of the knot.

\begin{figure*}[!t]
  \centering \includegraphics[width=18.0cm]{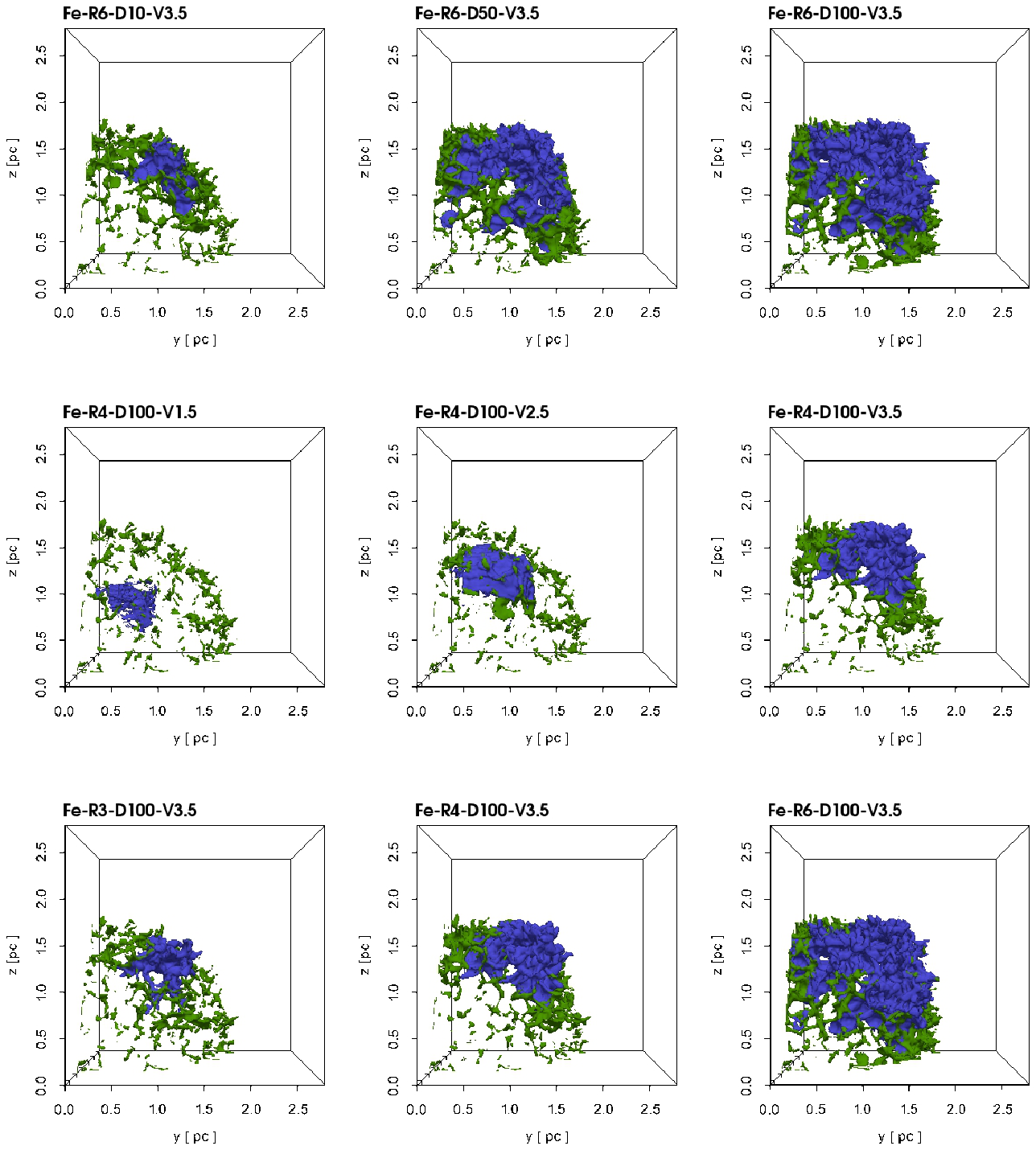}
  \caption{Spatial distributions of shocked Fe (blue) and Si/S
  (green) at $t = 340$~yr, in the case of a shrapnel initially
  located within the iron core (see models with $D_{\rm knot} =
  0.15 R_{\rm SNR}$ in Table~\ref{tab_knot}) and assuming a $20
  M_{\odot}$ progenitor MS star. The figure shows one octant of the
  domain (the SN explosion occurred at the origin of the 3D Cartesian
  coordinate system $(x_0, y_0, z_0) = (0, 0, 0)$) and the colored
  regions mark material with particle number density $n> 0.03$~cm$^{-3}$.
  The upper panels show the distributions for different initial
  density contrasts $\chi_{\rm n}$, the middle panels for different
  initial velocity contrasts $\chi_{\rm v}$, and the lower panels
  for different initial radii of the shrapnel $r_{\rm knot}$.}
  \label{fig_clumps}
\end{figure*}

Fig.~\ref{fig_clumps} shows the spatial distributions of shocked
Fe (blue) and Si/S (green) at the age of Cas\,A, for different
parameters of a shrapnel initially located within the iron core
($D_{\rm knot} = 0.15\,R_{\rm SNR}$; see Table~\ref{tab_knot}).
Depending on its initial density and velocity contrasts, the shrapnel
can produce a spatial inversion of ejecta layers, leading to the
Si/S-rich ejecta physically interior to the Fe-rich ejecta. A similar
effect is observed in Cas\,A as, for example, in the so-called SE
Fe piston: X-ray observations show that Fe-rich ejecta are at a
greater radius than Si/S-rich ejecta, and this has been interpreted
as an overturning of ejecta layers during the SN explosion (e.g.
\citealt{2000ApJ...528L.109H}). In our simulations, the spatial
inversion occurs because the piston is subject to hydrodynamic
instabilities which lead to some overturning of the layers in a way
analogous that that described by \cite{2006A&A...453..661K} for the
instabilities developed during a SN explosion. The inversion is evident
at $t = 340$~yr if the initial density contrast of the shrapnel was
$\chi_{\rm n} > 10$ and its velocity contrast was $\chi_{\rm v} >
2.5$ (see upper and middle panels of Fig.~\ref{fig_clumps}). Also
we found that the presence of the inversion does not depend on the
shrapnel size $r_{\rm knot}$, at least in the range of values
explored (see lower panel of Fig.~\ref{fig_clumps}). This result
puts a lower limit to both the density and velocity contrasts of
an overdense knot formed in the iron core few hours after the SN
explosion in order to produce the spatial inversion of Fe-rich
ejecta with Si-rich ejecta observed in Cas\,A.

\begin{figure*}[!t]
  \centering \includegraphics[width=14cm]{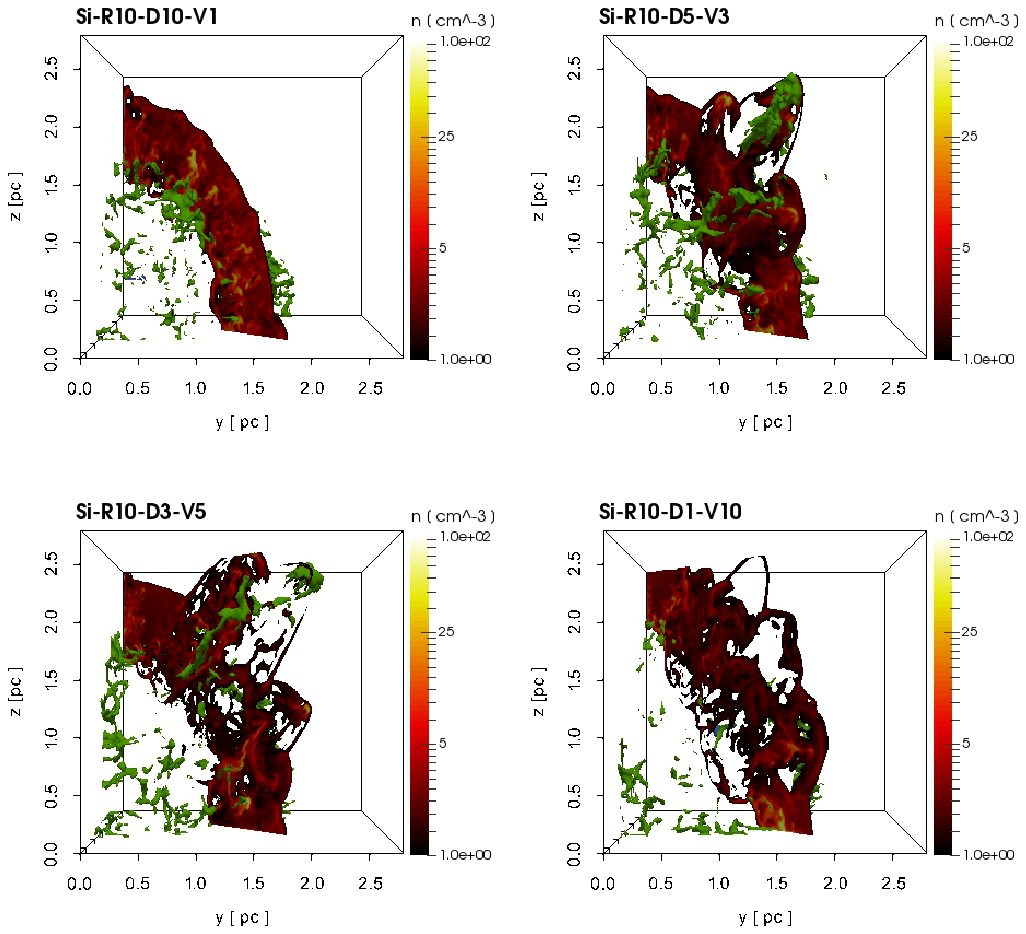}
  \caption{As in Fig.~\ref{fig_clumps} for a shrapnel initially
  located outside the iron core (see models with $D_{\rm knot} =
  0.35 R_{\rm SNR}$ in Table~\ref{tab_knot}). In this case, no
  significant shocked Fe is found. A slice showing the distribution
  of density, in log scale, in the intershock region is superimposed
  (see color table on the right of each panel). Each panel shows
  the result for different combinations of $\chi_{\rm n}$ and
  $\chi_{\rm v}$.} \label{fig_jets}
\end{figure*}

The final distribution of shocked Si/S in the case of a knot initially
located just outside the iron core is shown in Fig.~\ref{fig_jets}.
In this case, the simulations showed that knots with $\chi_{\rm
v}=1$ do not reach the remnant outline at $t=340$~yr even if their
density contrast $\chi_{\rm n}$ is up to 10 (upper left panel in
Fig.~\ref{fig_jets}); knots with $\chi_{\rm n}=1$ can perturb the
remnant outline at $t=340$~yr if $\chi_{\rm v}=10$ but without
producing a collimated Si/S-rich protrusion (see lower right panel
in Fig.~\ref{fig_jets}). Indeed a Si/S-rich shrapnel can reach the
remnant outline and protrude it, if the initial knot was both denser
and faster than the surrounding medium. Knots with
initial density and velocity contrasts larger than 1 produce Si-rich
jet-like features (see upper right and lower left panels in
Fig.~\ref{fig_jets}) which are similar to the NE and SW Si-rich
jets observed in Cas\,A.

\begin{figure}[!t]
  \centering \includegraphics[width=8.5cm]{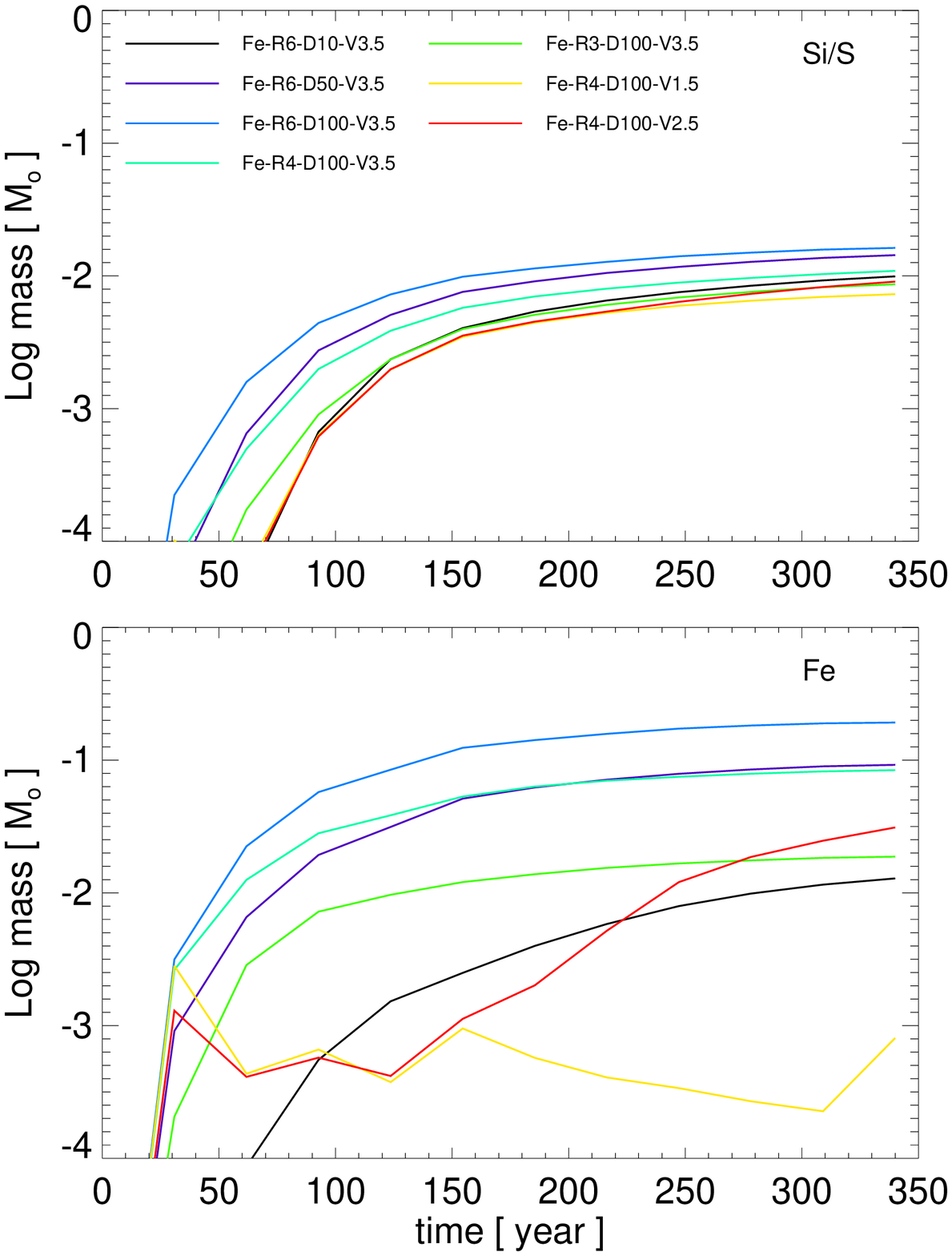}
  \caption{Amount of shocked Si/S (upper panel) and Fe (lower
  panel) vs time for models assuming a shrapnel initially located
  within the iron core ($D_{\rm knot} = 0.15 R_{\rm SNR}$) and
  characterized by different parameters (see Table~\ref{tab_knot}).}
  \label{evol_clumps}
\end{figure}

\begin{figure}[!t]
  \centering \includegraphics[width=8.5cm]{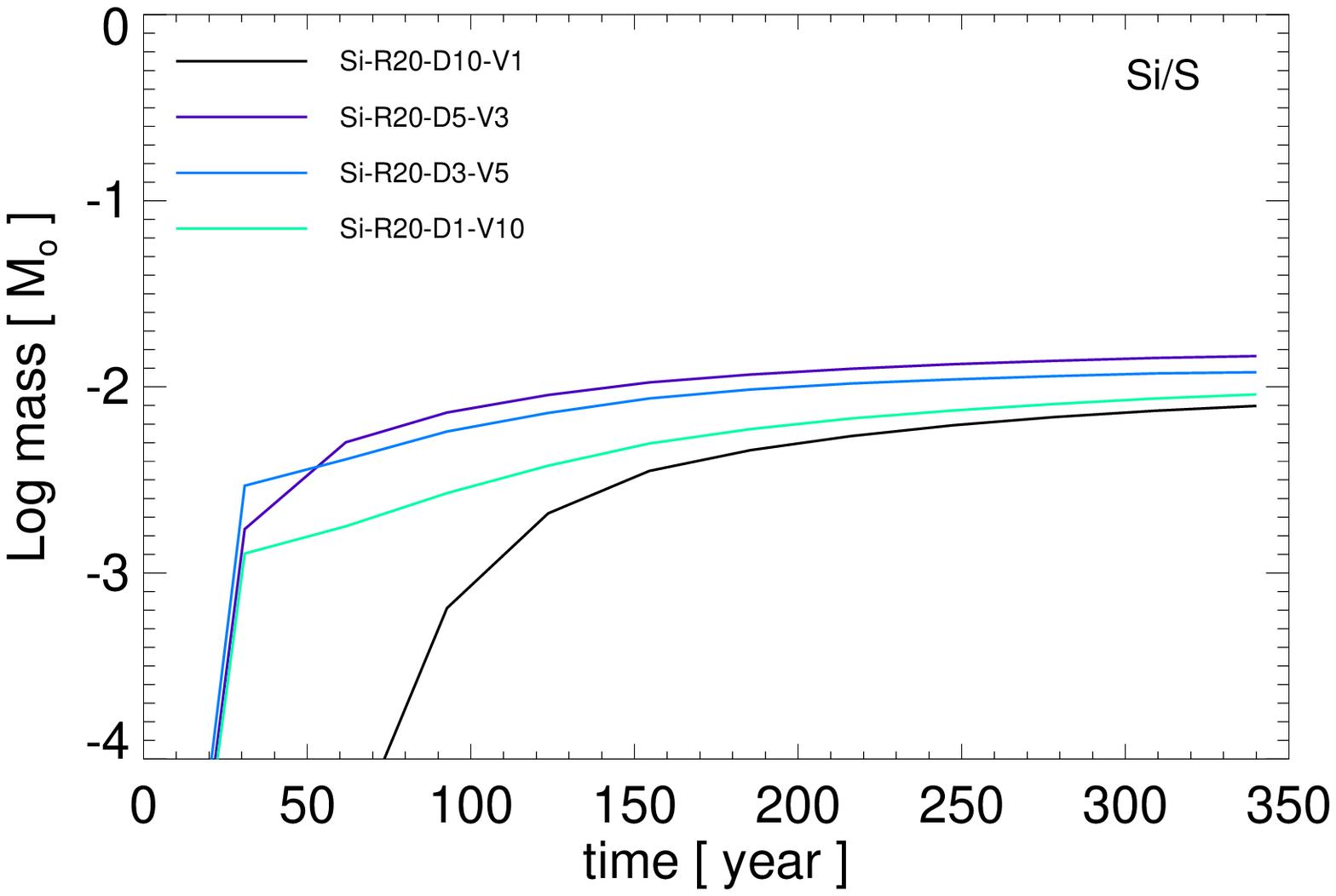}
  \caption{As in the upper panel of Fig.~\ref{evol_clumps} for
  models assuming a shrapnel initially located outside the iron
  core ($D_{\rm knot} = 0.35 R_{\rm SNR}$; see Table~\ref{tab_knot}).}
  \label{evol_jets}
\end{figure}

From the simulations, we derived the amount of shocked Fe and Si/S
for the different parameters of the shrapnel (see
Figs.~\ref{evol_clumps}-\ref{evol_jets}). We found that the amount
of shocked Si/S at $t=340$~yr is poorly affected by the initial
shrapnel parameters (within the ranges explored) and even by the
initial location of the knot (either inside or outside the iron
core). The final amount of shocked Si/S ranges between 0.008 and
$0.016 M_{\odot}$. This is mainly due to the fact that a significant
part of the Si/S shell interacts with the reverse shock even without
any initial shrapnel, so that the contribution of the latter to the
amount of shocked Si/S is poorly relevant. On the other hand, the
mass of shocked Fe at $t=340$~yr strongly depends on the initial
parameters of the shrapnel. In fact, in the absence of initial
anisotropies, the iron core is not reached by the reverse shock
during the first 340~yr of evolution (see Sect.~\ref{mod_spher}).
The shocked Fe observed in the simulations therefore is entirely
due to the shrapnel. We found that knots with $\chi_{\rm n} > 10$,
$\chi_{\rm v} > 2.5$, and $r_{\rm knot} > 0.03 R_{\rm SNR}$ lead
up to $\approx 0.1M_{\odot}$ of shocked Fe (see lower panel in
Fig.~\ref{evol_clumps}). Finally it is worth noting that significant
shocked Fe is produced only if the initial anisotropy was located
within the iron core.

\newpage
\subsection{Spatial distribution and chemical composition of
the Cas\,A ejecta}

Based on the results obtained in Sect.~\ref{sing_anis}, we searched
for the initial anisotropies that best reproduce the spatial
distribution of Fe and Si/S observed today in Cas\,A (e.g.
\citealt{2010ApJ...725.2038D, 2012ApJ...746..130H, 2013ApJ...772..134M}).
As reference model, we considered the case of a progenitor MS star
with $15\,M_{\odot}$ and injection rate $\eta = 10^{-4}$ (run
CAS-15MS-1ETA). The mass of ejecta, energy of explosion, radius of
progenitor star, and density of the RSG wind are those of run
SN-4M-2.3E-1ETA (see Table~\ref{tabmod}). Then we explored the space
of parameters characterizing the anisotropies, by adopting an
iterative process of trial and error to converge on parameters that
reproduce the spatial distribution and mass of shocked Fe and Si/S
inferred from observations (\citealt{2012ApJ...746..130H}).
Table~\ref{tab_anis} reports our best-fit parameters describing the
initial anisotropies.

\begin{table}
\caption{Parameters of the post-explosion anisotropies of ejecta describing the final morphology of Cas\,A}
\label{tab_anis}
\begin{center}
\begin{tabular}{lcccccc}
\hline
\hline
piston/jet &  $D_{\rm knot}$   &  $r_{\rm knot}$   &  $\chi_{\rm n}$  &  $\chi_{\rm v}$ & $M_{\rm knot}$ & $E_{\rm knot}$ \\
           &  [$R_{\rm SNR}$]  &  [$R_{\rm SNR}$]  &                  &                 & [$M_{\odot}$]  & [$10^{49} erg$] \\
\hline
Fe-rich SE &  0.15  &  0.05  &  100  &  4.2  & 0.10   & 5.0 \\
Fe-rich SW &  0.15  &  0.02  &  50   &  4.2  & 0.0015 & 0.076\\
Fe-rich NW &  0.15  &  0.06  &  50   &  4.2  & 0.10   & 4.8\\
Si-rich NE &  0.35  &  0.1   &  5.0  &  3.0  & 0.040  & 4.2 \\
Si-rich SW &  0.35  &  0.1   &  1.2  &  3.0  & 0.0091 & 1.0 \\
\hline
\end{tabular}
\end{center}
\end{table}

\begin{figure*}[!t]
  \centering \includegraphics[width=17.0cm]{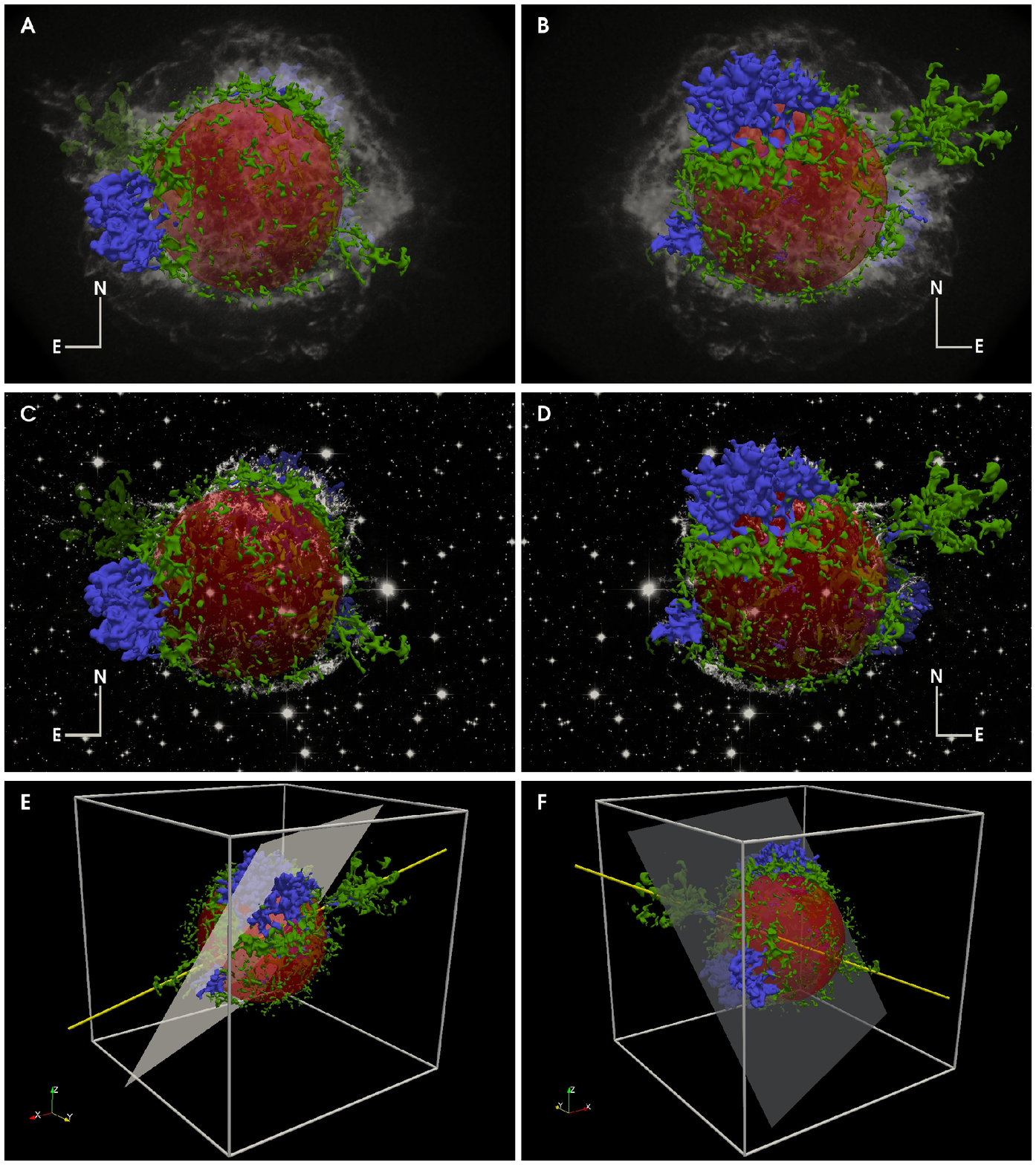}
  \caption{3D spatial distribution of Cas\,A shocked Fe (blue) and
  Si/S (green) derived from run CAS-15MS-1ETA (see Table~\ref{tab_elem}).
  Panels A and C show the 3D distribution assuming the vantage point
  at Earth (the perspective is in the plane of the sky); panels B
  and D show the same perspective but with the vantage point from
  behind Cas\,A (namely the perspective is rotated by $180^o$ about
  the northsouth axis); panels E and F show the distribution from
  arbitrary points of view. The transparent image in the upper
  panels (A and B) is a {\it Chandra} observation showing the hot
  shocked plasma in the wavelength band $[0.3, 10]$~keV (retrieved
  from www.nasa.gov); the transparent image in the middle panels
  (C and D) is a composite Hubble Space Telescope (HST) image
  sensitive to emission in cold O and S lines (retrieved from
  www.spacetelescope.org). The transparent red sphere marks the
  fiducial reverse shock; the transparent plane in panels E and F
  represents the plane where the initial Fe-rich anisotropies lie;
  the yellow line in panels E and F shows the direction of propagation
  of the Si-rich jet and counterjet. Refer to Movie 1 for an
  animation of these data.}
\label{ejecta_sh} \end{figure*}

Three of them are located within the iron core at $D_{\rm knot} =
0.15\,R_{\rm SNR}$, roughly lying in a plane oriented with an
$\approx -30^o$ rotation about the $x$-axis (namely the eastwest axis in
the plane of the sky) and an $\approx 25^o$ rotation about the
$z$-axis (the northsouth axis in the plane of the sky; see
Fig.~\ref{ejecta_sh}).
The other two are located just outside the iron core at $D_{\rm
knot} = 0.35\,R_{\rm SNR}$ on a line oriented with an $\approx
-67^o$ rotation about the $x$-axis and an $\approx 67^o$ rotation
about the $z$-axis (see Fig.~\ref{ejecta_sh}). The first three knots
reproduce the Fe-rich regions and the others the Si-rich NE jet and
SW counterjet observed today in Cas\,A. Table~\ref{tab_anis} reports
also the masses and energies of the pistons responsible for the
observed distribution of Fe and Si/S. We note that the total
kinetic energy of all the pistons is about $1.5\times 10^{50}$~erg.
Thus the pistons/jets represent a relatively small fraction ($\approx
7$\%) of the remnant's energy budget, at odds with previous estimates
(e.g. \citealt{2003A&A...398.1021W, 2006ApJ...644..260L}).

We performed additional simulations to explore also the case of a
progenitor MS star with $20\,M_{\odot}$ (run CAS-20MS-1ETA) and the
case in which there is no feedback of accelerated CRs ($\eta = 0$;
run CAS-15MS-0ETA). However, from the analysis of the distribution
of Si, S, and Fe, we did not find any appreciable difference of these
cases with our reference model (run CAS-15MS-1ETA). In the following,
therefore, we discuss in detail only the results of run CAS-15MS-1ETA,
mentioning the differences (if any) with the other two cases.

\subsubsection{Shocked ejecta} 

\begin{table}
\caption{Element Masses Derived from Models and Comparison with
{\it Chandra} Observations (\citealt{2012ApJ...746..130H})}
\label{tab_elem}
\begin{center}
\begin{tabular}{llcc}
\hline
\hline
Model            &  Element  &  $M_{\rm el,sh}/M_{\odot}$  &  $M_{\rm el,sh}/M_{\rm ej}$  \\
\hline
CAS-15MS-1ETA    &  Si/S              &  0.076    &  0.019  \\
                 &  Fe                &  0.19     &  0.048  \\
                 &  $M_{\rm ej, sh}$  &  3.66     &  0.92   \\
                 &  $M_{\rm ej}$      &  4        &  1      \\
CAS-15MS-0ETA    &  Si/S              &  0.080    &  0.020  \\
                 &  Fe                &  0.19     &  0.048  \\
                 &  $M_{\rm ej, sh}$  &  3.70     &  0.92   \\
                 &  $M_{\rm ej}$      &  4        &  1      \\
CAS-20MS-1ETA    &  Si/S              &  0.080    &  0.020  \\
                 &  Fe                &  0.18     &  0.045  \\
                 &  $M_{\rm ej, sh}$  &  3.66     &  0.92   \\
                 &  $M_{\rm ej}$      &  4        &  1      \\
\hline
Observations$^a$ &  Si/S              &  0.08     &  0.03   \\
                 &  Fe                &  0.14     &  0.04   \\
                 &  $M_{\rm ej, sh}$  &  2.84     &  0.90   \\
                 &  $M_{\rm ej}$      &  3.14     &  1      \\
\hline
\end{tabular}
\flushleft{$^a$ Values inferred from the analysis of {\it Chandra}
observations (\citealt{2012ApJ...746..130H}).}
\end{center}
\end{table}

In our favored model, the initial ejecta distribution (at day
$\approx 1$ since the SN event) is characterized by five large-scale
spherical knots (in addition to the smaller-scale clumpy structure).
From our model, we derived the total mass of shocked ejecta at
$t=340$~yr, $M_{\rm ej, sh} \approx 3.66\,M_{\odot}$ (see
Table~\ref{tab_elem}). This value is larger than that inferred by
\cite{2012ApJ...746..130H} from the analysis of {\it Chandra}
observations of Cas\,A, namely $M_{\rm ej, sh} \approx 2.84\,M_{\odot}$.
The latter estimate, however, has been derived by matching the
Cas\,A observations with 1D hydrodynamic models of SNR evolution.
We note that their best-fit model considers a total ejecta mass,
$M_{\rm ej} \approx 3.14 M_{\odot}$, lower than that adopted in our
simulations ($M_{\rm ej} \approx 4 M_{\odot}$). Indeed, considering
the fraction of the total ejecta mass that is shocked at $t=340$~yr,
$M_{\rm ej, sh}/M_{\rm ej}$, we found that the value predicted with
our model ($\approx 92$\,\%) is in agreement with that derived by
\cite{2012ApJ...746..130H} ($\approx 90$\,\%).

Fig.~\ref{ejecta_sh} shows the 3D spatial distribution of shocked
Fe and Si/S derived from run CAS-15MS-1ETA, considering different
points of view. An on-line animation has been provided that shows
the 3D distribution rotated completely about the northsouth axis
(Movie 1). The ejecta morphology is dominated by the three Fe-rich
regions caused by the initial anisotropies located in the iron core
(see Table~\ref{tab_anis}). We note that the SE Fe-rich region has
more a jet-like structure if compared with the other regions in
agreement with the observations (\citealt{2010ApJ...725.2038D}).
This is due to the higher density contrast of the corresponding
initial anisotropy (see Table~\ref{tab_anis}). As expected on the
base of the results of Sect.~\ref{sing_anis}, the initial anisotropies
produce a spatial inversion of ejecta layers at the age of Cas\,A,
leading locally to Fe-rich ejecta placed at a greater radius than
Si/S-rich ejecta.

A striking aspect is that the Fe-rich regions are circled by rings
of Si/S-rich ejecta. In particular we can identify clearly two
complete rings around the NW and SE Fe-rich regions (see Movie 1).
These features resemble the cellular structure of [Ar II], [Ne II],
and Si XIII observed in Cas\,A that appears as rings on the surface
of a sphere (e.g. \citealt{2010ApJ...725.2038D}).  The Si/S-rich
rings form as a result of the high velocity Fe-rich ejecta pistons
(the shrapnels) in a way similar to that suggested by
\cite{2010ApJ...725.2038D}.  Each piston pushes out the chemically
distinct layers above and, at some point, it breaks through some
of them, leading to the spatial inversion of ejecta layers (see
Sect.~\ref{sing_anis}). Then the material of the outer layers (in
particular the Si/S) is swept out by the piston and progressively
accumulates at its sides. As a result, when the piston encounters
the reverse shock, a region of shock-heated Fe forms which is
enclosed by a ring of shock-heated Si/S.

The post-explosion anisotropies located just outside the iron core
are responsible for the Si/S-rich jets (see Fig.~\ref{ejecta_sh}
and Movie 1). In this case the shrapnels (given the initial values
of their density and velocity contrasts) break through the outer
ejecta layers, thus not preserving locally the original onion-skin
nucleosynthetic layering. Then the shrapnels are able to protrude
the remnant outline, thus forming wide-angle ($\approx 40^o$)
opposing streams of Si/S-rich ejecta in the NE and SW quadrants.
This is consistent with the scenario proposed to explain the origin
of the Si/S-rich NE jet and SW counterjet observed in Cas\,A (e.g.
\citealt{2006ApJ...645..283F}). The streams travel at velocities
up to $\approx 10000$~km~s$^{-1}$ in agreement with the values
inferred from the observations (\citealt{2001ApJS..133..161F,
2004ApJ...615L.117H, 2015arXiv151205049F}).

\begin{figure}[!t]
  \centering \includegraphics[width=8.5cm]{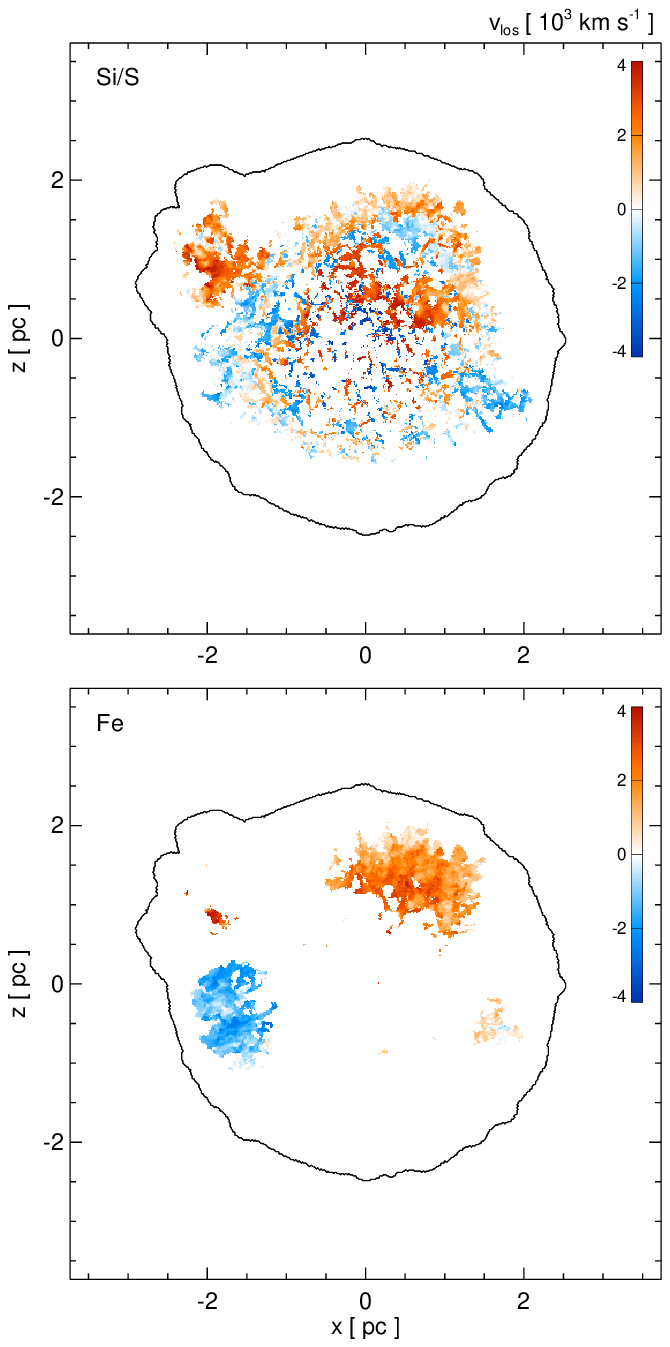}
  \caption{2D maps of average emission-measure-weighted velocity
  along the line-of-sight, $v_{los}$, for shocked Si/S (upper panel)
  and Fe (lower panel) derived from run CAS-15MS-1ETA. The approximate
  velocity range is $\pm 4000$~km~s$^{-1}$ (see color bar on the
  right in units of $10^3$~km~s$^{-1}$). These images correspond
  to effective Doppler maps derived from observations.} \label{fig_vel}
\end{figure}

The pistons correspond to regions where relatively faster-moving
ejecta were expelled. Fig.~\ref{fig_vel} shows the average
emission-measure-weighted velocity of shocked Fe and Si/S along the
line of sight derived from our model. The shocked Si/S is mostly
concentrated in the redshifted jet to the NE and a large redshifted
ring-like feature to the NW. The latter is the result of the piston
which breaks through the Si/S layer and is responsible for the large
Fe-rich region to the NW. These two features are those with the
highest absolute values of velocity along the line of sight, $\approx
4000$~km~s$^{-1}$. In the case of the NE jet (inclined with an angle
of $23^o$ with respect to the plane of the sky), this corresponds
to the high velocity values of the streams of $\approx 10000$~km~s$^{-1}$
derived from the simulation. The SW jet and the SE piston are visible
as blueshifted material. The shocked Fe is concentrated in the most
prominent redshifted area to the NW and the blueshifted area to the
SE. We note that the velocity pattern of shocked Fe is remarkably
similar to the Doppler images derived from observations of Cas\,A
(e.g. \citealt{2002A&A...381.1039W, 2010ApJ...725.2038D}). In
particular it matches the approximate velocity range inferred from
observations, namely $\pm 4000$~km~s$^{-1}$.

From the models we calculated the masses of Fe and Si/S in shocked
ejecta at $t=340$~yr (see Table~\ref{tab_elem}) and compared them
with the values inferred from the analysis of {\it Chandra} observations
(\citealt{2012ApJ...746..130H}). We found that the models adopting
the parameters of the anisotropies summarized in Table~\ref{tab_anis}
reproduce quite well the mass of shocked Fe\footnote{The modeled
values are slightly higher than those observed, but this depends
on the fine tuning of the parameters of the asymmetries.}. On the
other hand, all our models slightly underestimate the fraction of
mass of shocked Si/S, $M_{\rm Si/S,sh}/M_{\rm ej}$, by $\approx
30$\% (see Table~\ref{tab_anis}). This was somehow expected on the
basis of the results of Sect.~\ref{sing_anis}. In fact,
the effect of the initial asymmetries is only to slightly increase the
mass of shocked Si/S with respect to the case of a spherically symmetric
explosion. Thus a way to increase significantly the mass of shocked
Si/S might be to change the initial isotopic composition adopted
for the ejecta. However, we found that our result does not change
either if we adopt a progenitor MS star with $20\,M_{\odot}$ (run
CAS-20MS-1ETA in Table~\ref{tab_elem}) or if we neglect the effects
of CRs acceleration (run CAS-15MS-0ETA). Nevertheless we note that
our models as well as the values inferred from the observations are
subject to some uncertainties, so that a discrepancy of the order
of 30\% may be considered satisfactory.

\begin{figure}[!t]
  \centering \includegraphics[width=8.5cm]{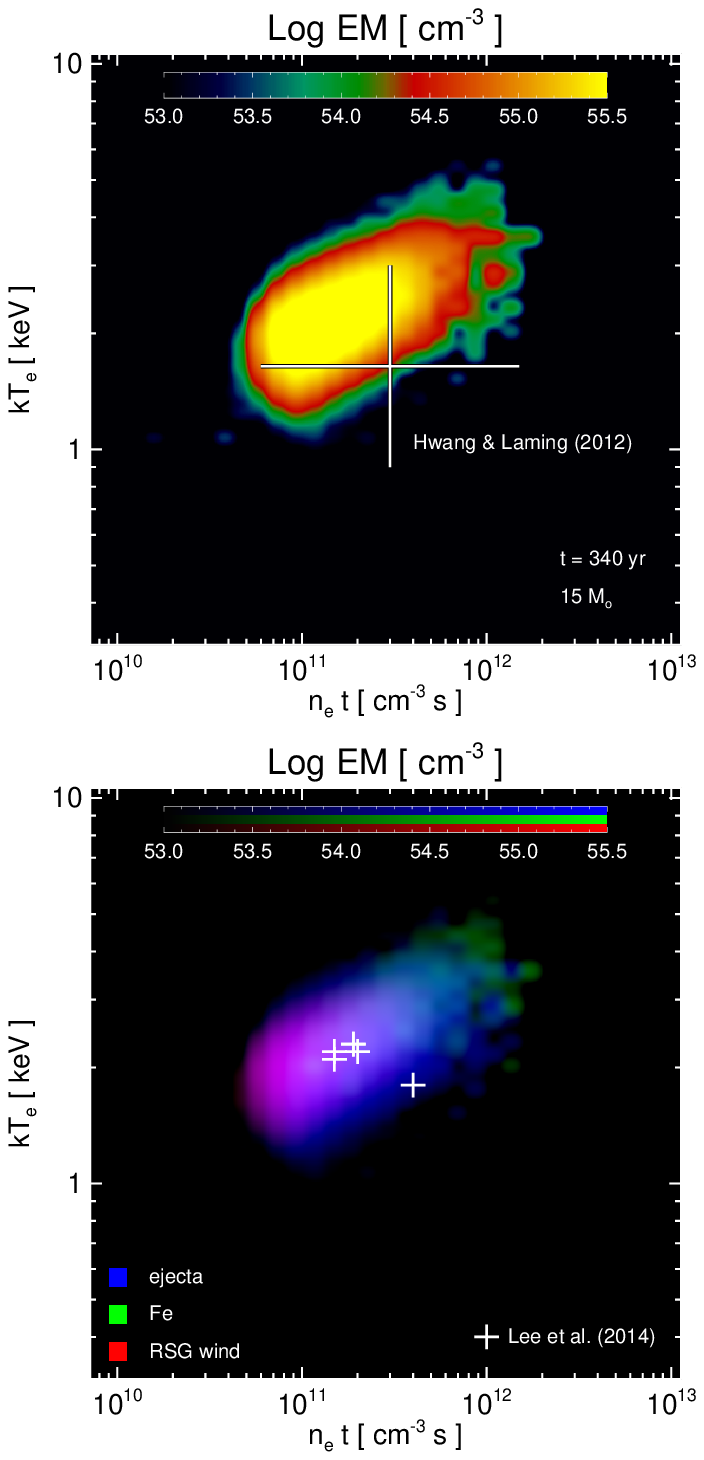}
  \caption{Upper panel: distribution of emission measure vs. electron
  temperature $kT_{\rm e}$ and ionization age $n_{\rm e} t$ at $t=
  340$~yr derived from run CAS-15MS-1ETA. The white cross shows the
  ranges of $kT_{\rm e}$ and $n_{\rm e} t$ values inferred from
  {\it Chandra} observations (\citealt{2012ApJ...746..130H}). Lower
  panel: corresponding three-color composite image of the emission
  measure distribution. The colors show the contribution to emission
  measure from the shocked ejecta (blue), shocked Fe (green), and
  shocked RSG wind (red). The white crosses mark the values derived
  by \cite{2014ApJ...789....7L} from the analysis of regions dominated
  by thermal emission of shocked ambient gas.}
  \label{tau_te}
\end{figure}

Fig.~\ref{tau_te} shows the emission measure distribution as a
function of electron temperature, $kT_{\rm e}$, and ionization age,
$n_{\rm e} t$, for the shocked plasma at the age of Cas\,A. The
X-ray emitting plasma is largely out of equilibrium of ionization
with the emission measure distribution peaking at $kT_{\rm e} \approx
2$~keV and $n_{\rm e} t \approx 10^{11}$~cm$^{-3}$~s in a region
dominated by shocked ISM (see upper panel in Fig.~\ref{tau_te}).
These values are in excellent agreement with the best-fit parameters
derived by \cite{2014ApJ...789....7L} from the analysis of X-ray
spectra extracted from several regions around the outermost boundary
dominated by thermal emission of shocked ambient gas (see lower
panel in Fig.~\ref{tau_te}). A more complete comparison of our model
results with observations is obtained by considering the analysis
of {\it Chandra} observations of \cite{2012ApJ...746..130H}.  These
authors provided a consistent spectral characterization of the
entire remnant of Cas\,A by defining a grid of macropixels across
the remnant and analyzing every spectrum in the grid. In such a
way, they were able to derive the distributions of ionization ages
and electron temperatures of the entire remnant. We found that our
simulation predicts $kT_{\rm e}$ and $n_{\rm e} t$ values in the
observed ranges. The distributions of $kT_{\rm e}$ and $n_{\rm e}
t$ derived from the simulation are highly peaked in agreement with
the findings of \cite{2012ApJ...746..130H} who suggested that the
peaked distributions result from the multiple secondary shocks
following reverse shock interaction with ejecta inhomogeneities.
Fig.~\ref{tau_te} shows also that the shocked Fe is at an advanced
ionization age ($n_{\rm e} t\approx 10^{12}$~cm$^{-3}$~s) relative
to the other elements.  This is also in nice agreement with the
observations (e.g.  \citealt{2003ApJ...597..362H, 2012ApJ...746..130H}).

\subsubsection{Unshocked ejecta} 

In our model, the total mass of unshocked ejecta at the age of
Cas\,A is $\approx 0.34\,M_{\odot}$. This value is in good
agreement with that inferred from the analysis of low frequency
($<100$~MHz) radio observations ($\approx 0.39\,M_{\odot}$;
\citealt{2014ApJ...785....7D}) and that derived by
\cite{2012ApJ...746..130H} by interpreting the {\it Chandra}
observations with hydrodynamic models ($\approx 0.30\,M_{\odot}$).

\begin{figure*}[!t]
  \centering \includegraphics[width=17.0cm]{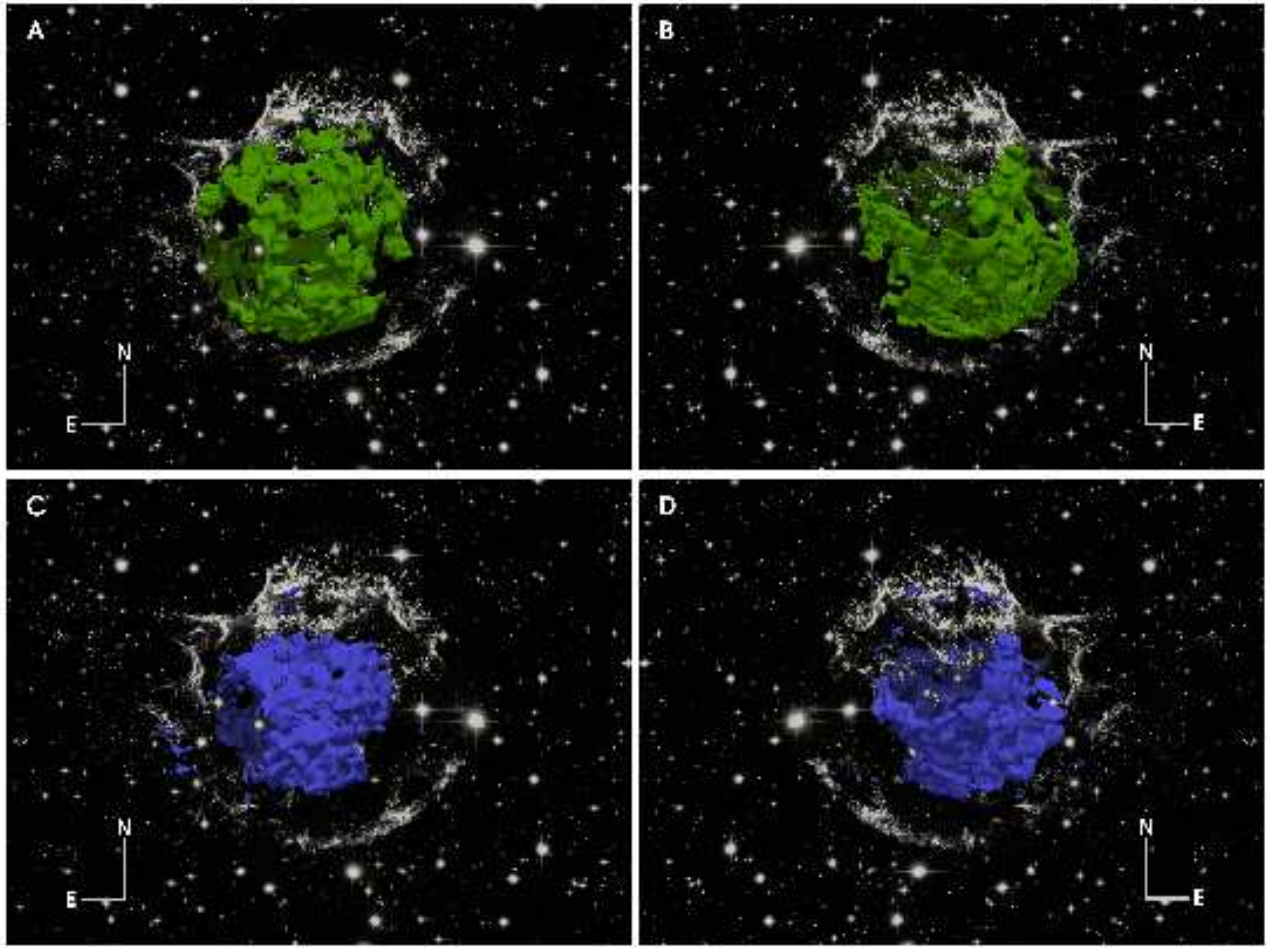}
  \caption{3D spatial distribution of Cas\,A unshocked Si/S (green,
  upper panels) and Fe (blue, lower panels) derived from run
  CAS-15MS-1ETA. Panels A and C show the 3D distribution assuming
  the vantage point at Earth; panels B and D show the same perspective
  but with the vantage point from behind Cas\,A. As in Fig.\ref{ejecta_sh},
  the transparent image in the panels is a composite 
  HST image. Refer to Movie 2 for an animation of these
  data.} \label{ejecta_unsh}
\end{figure*}

The 3D spatial distributions of unshocked Fe and Si/S are reported
in Fig.~\ref{ejecta_unsh}. An on-line animation shows the 3D
distribution rotated completely about the northsouth axis (Movie
2). The majority of the unshocked Si/S follows roughly the original
onion-skin nucleosynthetic layering. However, the original Si/S
layer is characterized by five large cavities corresponding to the
directions of propagation of the post-explosion anisotropies
(pistons/jets). The regions of shocked Fe observed in the
main-shell and the NE and SW jets are located exactly above the
cavities. We also found that the cavities are physically connected
to some of the rings of shocked Si/S in the main-shell as, for
instance, in the case of the NW and SE cavities (compare
Figs.~\ref{ejecta_sh} and \ref{ejecta_unsh}, and Movies 1 and 2).
The average structure of the unshocked Si/S is somehow reminiscent
of the bubble-like morphology characterizing the remnant interior
of Cas\,A, inferred from the analysis of near-infrared spectra of
the remnant including the [S III] 906.9 and 953.1 nm
(\citealt{2015Sci...347..526M}).

The Fe-rich NW and SE pistons are responsible for the largest
cavities in our simulations, with the NW cavity with a radius of
$\approx 0.88$~pc and the SE cavity with a radius of $\approx
0.54$~pc. These values are also in good agreement with those estimated
from the analysis of observations (\citealt{2015Sci...347..526M}).
On the other hand, the distribution of unshocked S derived from the
analysis of near-infrared spectra appears much more structured than
in our simulations. This may suggest that a number of smaller-scale
pistons with lower velocity and/or density contrast (not considered
in our simulations) might be present in Cas\,A.

The structure of Si/S is filled by unshocked Fe (see Fig.~\ref{ejecta_unsh}
and Movie 2) with a density below 0.1~cm$^{-3}$. Most of the Fe is
concentrated in the remnant core and shows a clumpy structure in
agreement with the expectation of strong instabilities associated
with the explosion reverse shock during the first few hours after
the SN explosion. The distribution of Fe is roughly spherically
symmetric, although large cavities are evident which correspond to
the regions of the initial pistons. The total mass of unshocked Fe
derived from the model is $\approx 0.1\,M_{\odot}$.

Our model predicts a low-density and low-temperature environment
in the unshocked ejecta. The average density derived from the
simulation is $\rho \approx 2\times 10^{-24}$~g~cm$^{-3}$. From the
analysis of radio observations of Cas\,A, \cite{2014ApJ...785....7D}
derive a density $\rho \approx 5.5 \times 10^{-24}$~g~cm$^{-3}$,
assuming a uniform density ejecta distribution throughout the remnant
interior with no clumping. The hydrodynamic models considered by
\cite{2012ApJ...746..130H} to interpret the {\it Chandra} observations
of Cas\,A predict a density $\rho \approx 1\times 10^{-24}$~g~cm$^{-3}$
at an age of 330 yr. The temperature of the unshocked ejecta is
rather low due to the rapid expansion of the SNR and ranges between
$\approx 10$~K (close to the origin of the SN explosion) and $\approx
10^3$~K (immediately before the reverse shock). These temperatures
are consistent with the analysis of the {\it Spitzer} infrared data
(e.g.  \citealt{2009ApJ...697...29E}) and with the evidence that
unshocked dust temperatures of about 35~K are observed in the
infrared band (e.g. \citealt{2010ApJ...713..356N}).

\section{Summary and conclusions}
\label{sec4}

We studied the evolution of ejecta in the SNR Cas\,A with the aim
to investigate the origin of the asymmetries observed today in its
morphology. In particular, we investigated the scenario of high
velocity pistons of ejecta emerging from the SN explosion
proposed by \cite{2010ApJ...725.2038D} to explain the distribution
of Fe-rich regions and Si-rich jets observed today in Cas\,A.  To
this end, we developed a hydrodynamic model describing the evolution
of Cas\,A from the immediate aftermath of the SN explosion to the
remnant expansion through the wind of the progenitor RSG,
thus covering the evolution of the system from few seconds after
the SN event till the current age of the remnant ($t=340$~yr). The
model includes the effects on shock dynamics due to back-reaction
of accelerated cosmic rays and describes the initial structure of
the ejecta through small-scale clumping of material and larger-scale
anisotropies, according to indications from theoretical studies
(e.g. \citealt{1993ApJ...419..824L, 2002ApJ...579..671W,
2006A&A...453..661K, 2008ARA&A..46..433W, 2010A&A...521A..38G}, and
references therein). The model follows the evolution of the
post-explosion isotopic composition of the ejecta in order to trace
the distribution of Si, S, and Fe, namely the elements characterizing
most of the anisotropies (e.g. jets, pistons) identified in the
morphology of Cas\,A (e.g. \citealt{2010ApJ...725.2038D,
2013ApJ...772..134M}).

We explored the parameter space of the model searching for the values
of ejecta mass, $M_{\rm ej}$, and explosion energy, $E_{\rm SN}$, best
reproducing the radii and velocities of the forward and reverse shocks
as observed at the current time ($t=340$~yr), and the density of the
shocked RSG wind inferred from observations. The best match was found
for models with $M_{\rm ej} \approx 4 M_{\odot}$ and $E_{\rm SN}\approx
2.3\times 10^{51}$~erg. These values are in good agreement with those
estimated from the analysis of observations (\citealt{2003ApJ...597..347L,
2003ApJ...597..362H, 2006ApJ...640..891Y}). It is worth noting
that, in our simulations, the envelope mass, $M_{\rm ej}$, was fixed
equal to $4 M_{\odot}$ (according to \citealt{2006ApJ...640..891Y} and
\citealt{2009A&A...503..495V}) in order to reduce the computational cost
in the exploration of the parameter space. On the other hand, an envelope
mass of $\approx 3-5\, M_{\odot}$ would not considerably deviate from
the observations. We expect that, adopting lower values of $M_{\rm ej}$,
the explosion energy $E_{\rm SN}$ required to fit our observational
constraints would be larger than that found here\footnote{For instance,
\cite{2003ApJ...593L..23C} found an explosion energy $E_{\rm SN}
\approx 4\times 10^{51}$~erg with an ejecta mass of $M_{\rm ej}
= 3.2\,M_{\odot}$.}. The opposite is expected for higher values of
$M_{\rm ej}$. On the other hand, the explosion energy that we found for
$M_{\rm ej} \approx 4 M_{\odot}$ is very close to that inferred from the
observations ($\approx 2\times 10^{51}$~erg; \citealt{2003ApJ...597..347L,
2003ApJ...597..362H}), so that we are confident that the envelope mass
adopted here is not far away from the true value. Our best-fit models
also predict that the radius of the progenitor RSG at the explosion was
$R_0 \approx 300-400\,R_{\odot}$. It is interesting to note that this
value is compatible with a RSG that has lost a significant fraction of its
H-rich envelope before the SN event (e.g. \citealt{2012ApJ...757...31B}).

The evidence of non-thermal emission in Cas\,A suggests that effective
acceleration of CRs occurs at the shock fronts. Thus we investigated
the effects of back-reaction of accelerated CRs on the shock dynamics
for different injection rates $\eta$. We found that the fraction
of explosion energy converted to CRs at the current epoch increases
with $\eta$ and ranges between $\approx 9\times 10^{49}$~erg ($\eta =
10^{-4}$) and $\approx 2.3\times 10^{50}$~erg ($\eta = 10^{-3}$).
These values are larger than those inferred from the observations
(namely $\approx 4\times 10^{49}$~erg; \citealt{2010ApJ...710L..92A,
2013ApJ...779..117Y}), suggesting that the injection rate in Cas\,A
is $\eta < 10^{-4}$. Since the plasma compressibility increases in the
presence of CR acceleration, a less dense RSG wind is needed to fit the
observed radii and velocities of the forward and reverse shocks. In the
case of a very efficient CR acceleration ($\eta = 10^{-3}$), our model
predicts a wind density $n_{\rm w} \sim 0.4$~cm$^{-3}$ at $r \approx
2.5$~pc, suggesting that the swept-up mass of the RSG wind is $M_{\rm
w,sh} \approx 3 M_{\odot}$ (see Eq.~\ref{wind_mass}). Note that the
value $M_{\rm w,sh}\approx 6 M_{\odot}$ derived in Sect.~\ref{prog_star}
for negligible CR acceleration is an upper limit to the shocked mass of
the wind.

It is interesting to note that models with different $\eta$
required the same explosion energy to fit our observational
constraints, despite the amount of energy lost in CRs acceleration
is larger for higher values of $\eta$. This is explained because a
lower density of the RSG wind was required to fit the observed
density of the shocked wind. In fact, on one hand, the energy lost
produces a slowdown and, on the other hand, the lower wind density
produces a speedup of the forward shock. We found that the two
effects cancel out each other in our simulations, so that the same
explosion energy is required for different values of $\eta$.

We investigated the effects that high-velocity pistons emerging
from the SN explosion have on the final remnant morphology. In
particular, we explored the parameter space of the initial pistons
with the aim to reproduce the spatial distribution and the masses
of shocked Fe and Si/S inferred from the observations of Cas\,A.
For this exploration, we considered the initial size of
the shrapnels and their contrasts of density and velocity with
respect to the surrounding ejecta. Since the physical quantities
characterizing the shrapnels are set relatively to the surrounding
medium, we do not expect significant changes in our results if
different total mass of ejecta (which affects their density) and/or
explosion energy (affecting the density and velocity of ejecta)
were adopted in the simulations. In all the cases explored we found
that the initial shrapnels are gradually fragmented into smaller
clumps as the remnant evolves.  In particular this happens when the
shrapnels are compressed and heated by the reverse shock arising
from the interaction of the remnant with the RSG wind.

The model best matching the observations predicts that, at least,
five large-scale anisotropies have developed in the immediate
aftermath of the SN explosion. Three of them were located within
the iron core and reproduce the Fe-rich regions observed today in
Cas\,A. The other two were located just outside the iron core and
reproduce the Si-rich NE jet and SW counterjet. The
parameters characterizing these anisotropies (namely their initial
size, and the velocity and density contrasts) are consistent with
those found through multi-dimensional modeling of SN explosions for
the dense knots produced by Rayleigh-Taylor instabilities seeded
by flow structures resulting from neutrino-driven convection (e.g.
\citealt{2003A&A...408..621K, 2012ApJ...755..160E}). We determined
the energies and masses of the initial anisotropies and found that
they had a total mass of $\approx 0.25\,M_{\odot}$ and a total
kinetic energy of $\approx 1.5\times 10^{50}$~erg, thus representing
a small fraction of the total ejecta mass ($\approx 5$\%) and of
the remnant's energy budget ($\approx 7$\%).

We note that our simulations predict that the initial
pistons were faster than the surrounding ejecta. Interestingly,
\cite{2015A&A...577A..48W}, through 3D modeling of core-collapse
SNe, have found that SN explosions from RSG progenitors with
$15\,M_{\odot}$ (at variance with SNe from blue supergiant progenitors)
present a large development of fast plumes of Ni-rich ejecta which
grow into extended fingers from which fast metal-rich clumps detach.
This leads to a global metal asymmetry characterized by pronounced
clumpiness, and a deep penetration of Ni fingers into the overlying
layers of ejecta.

In regions not affected by large-scale anisotropies, the average
chemical stratification of the ejecta few days after the SN event
is roughly maintained in the subsequent evolution until the current
age of Cas\,A. Thus, although the rapid ejecta expansion is certainly
not homologous because clumps and instabilities may easily develop
(e.g. \citealt{2001ApJ...549.1119W, 2012ApJ...749..156O}), our model
suggests that the abundance pattern observed in young SNRs keeps
memory of the radial distribution of heavy elements in the aftermath
of the explosion in regions not affected by the large-scale
anisotropies developed during the SN event. This result explains
the evidence that, apparently, the onion-skin nucleosynthetic
layering of the SN has been preserved in some regions of Cas\,A
(e.g.  \citealt{2006ApJ...645..283F, 2010ApJ...725.2038D}), namely
those not affected by the initial pistons.

On the other hand, the chemical stratification is not preserved in
regions strongly affected by the pistons/jets propagation. The pistons
responsible for Si/S-rich jets break through the outer ejecta layers,
protruding the remnant outline and forming opposing streams of Si/S-rich
ejecta in the NE and SW quadrants. The Fe-rich pistons produce a spatial
inversion of ejecta layers, leading to the Si/S-rich ejecta physically
interior to the Fe-rich ejecta. In fact, each piston is subject to
hydrodynamic instabilities during its propagation and this lead to
some overturning of the chemical layers. Again, this result matches
nicely with the evidence that Fe-rich ejecta are at a greater radius
than Si/S-rich ejecta in the Fe-rich regions of shocked ejecta observed
in Cas\,A (e.g.  \citealt{2000ApJ...528L.109H}).

A striking feature is that the regions of Fe-rich shocked ejecta are
circled by rings of Si/S-rich shocked ejecta. These rings are the
result of the dynamics of high-velocity pistons emerging from the
SN explosion. As a snowplow, each piston pushes the layers above
to the side, causing a progressive accumulation of chemically
distinct material (in particular Si/S) around the piston itself.
As a result, when the piston encounters the reverse shock, a central
region of shocked Fe circled by a ring enriched of shocked Si/S
forms. An example is the region of Fe-rich shocked ejecta to the
NW (see Fig.~\ref{ejecta_sh} and Movie 1).

The modeled rings are consistent with the bright rings of [Ar
II], [Ne II], and Si XIII around the Fe-rich regions observed in
Cas\,A (e.g. \citealt{2004ApJ...614..727M, 2007AJ....133..147P,
2010ApJ...725.2038D}). Thus our model provides evidence that
high-velocity pistons emerging from the SN explosion may explain
the origin of the bright rings observed in Cas\,A, supporting the
original scenario proposed by \cite{2010ApJ...725.2038D}. We note
however that the observed rings are much more pronounced than those
in our simulations. A role might be played by the magnetic field
which is neglected in our model. If the ejecta are magnetized, the
clumps are expected to be preserved and longer-living because the
field would envelope the ejecta clumps, limiting the growth of
hydrodynamic instabilities contributing to their fragmentation (e.g.
\citealt{2012ApJ...749..156O}). As a result, the clumps of chemically
distinct material pushed to the side of the pistons might survive
for a longer time, producing more evident and brighter rings.

A further support to the scenario of high-velocity pistons comes
from the analysis of the deviations from equilibrium of ionization
of shocked plasma. Our simulations predict that the shocked Fe is
at an advanced ionization age ($n_{\rm e} t\approx 10^{12}$~cm$^{-3}$~s)
relative to the other elements, in excellent agreement with {\it
Chandra} observations of Cas\,A (e.g. \citealt{2012ApJ...746..130H}).
It is interesting to note that the evidence of an advanced ionization
age of shocked Fe is considered as an argument against the origin
of Fe-rich regions from expanding plumes of radioactive $^{56}$Ni-rich
ejecta which predicts much lower ionization ages (e.g.
\citealt{1993ApJ...419..824L, 2001ApJ...557..782B, 2015Sci...347..526M}).

Finally we analyzed also the distribution of unshocked ejecta and
found that our model reproduces the main features observed in the
radio and near-infrared bands. The distribution of unshocked Si/S
is characterized by large cavities corresponding to the directions
of propagation of the pistons/jets. This explains why the cavities
observed in near-infrared observations are physically connected to
the bright rings in the main-shell (e.g. \citealt{2015Sci...347..526M}).
Our model predicts that the structure of Si/S is filled by low-density
unshocked Fe. We estimated a total mass of unshocked Fe of $\approx
0.1\,M_{\odot}$.

\acknowledgments
We thank the anonymous referee for useful suggestions that have
allowed us to improve the paper. This paper was partially funded
by the PRIN INAF 2014 grant ``Filling the gap between supernova
explosions and their remnants through magnetohydrodynamic modeling
and high performance computing''. MLP acknowledges financial support
from INAF-OAPA and CSFNSM. The software used in this work was, in
part, developed by the U.S. Department of Energy-supported Advanced
Simulation and Computing/Alliance Center for Astrophysical Thermonuclear
Flashes at the University of Chicago. We acknowledge that the results
of this research have been achieved using the PRACE Research
Infrastructure resource MareNostrum III based in Spain at the
Barcelona Supercomputing Center (PRACE Award N.2012060993).

\bibliographystyle{apj}
\bibliography{references}


\end{document}